\documentclass[journal,onecolumn]{IEEEtran}
\usepackage{todonotes}
\usepackage{research5IEEE}
\usepackage{mathrsfs,url,psfrag,graphicx,epsfig,mathtools,research5IEEE,amsmath,upgreek,
	amsfonts,amssymb,amsthm,tikz,array,wasysym,pifont,dsfont,balance,epstopdf}

\usetikzlibrary{arrows}

\IEEEoverridecommandlockouts
\newtheorem{remark}{Remark}
\newcommand{\outage}{\textnormal{outage}}

\newcommand{\GameNF}{\mathcal{G}(b) = \left(\mathcal{K}, \left\lbrace\mathcal{A}_k \right\rbrace_{k \in \mathcal{K}},\left\lbrace
u_{k}\right\rbrace_{ k \in \mathcal{K}}\right)}

\newcommand{\gameNF}{\mathcal{G}(b)}

\newcommand{\ds}{\displaystyle}

\newcommand{\GMAC}{ \textnormal{G-MAC}$(K,b)$ }

\newcommand{\SUD}{ \mathrm{SUD}}
\newcommand{\SIC}{ \mathrm{SIC}}
\newcommand{\coop}{ \mathrm{coop}}
\newcommand{\ind}{ \mathrm{ind}}

\newtheorem{lemma}{Lemma}
\newtheorem{theorem}{Theorem}

\newtheorem{definition}{Definition}
\newtheorem{proposition}{Proposition}

\begin{document}
\title{Decentralized Simultaneous Information and Energy Transmission in $K$-User Multiple Access Channels}
\author{
\IEEEauthorblockN{Selma Belhadj Amor, Samir M. Perlaza, and H. Vincent Poor}

\thanks{Selma Belhadj Amor was with the Institut National de Recherche en Informatique et en Automatique (INRIA) at Lyon, France and was visiting the Department of Electrical Engineering at Princeton University, Princeton, NJ 08544, USA. She is now with the Department of Electrical and Computer Engineering at the National University of Singapore, Singapore 119077 (email:elesba@nus.edu.sg).}

\thanks{Samir M. Perlaza is with the Institut National de Recherche en Informatique et en Automatique (INRIA) at Lyon, France (samir.perlaza@inria.fr). He is also with the Department of Electrical Engineering at Princeton University, Princeton, NJ 08544, USA.}

\thanks{H. Vincent Poor is with the Department of Electrical Engineering at Princeton University, Princeton, NJ 08544, USA (poor@princeton.edu).}

\thanks{Part of this work was presented at the 50th Annual
	Conference on Information Sciences and Systems (CISS), Princeton NJ, USA, March 2016. This research is supported in part by the European Commission under Marie Sk\l{}odowska-Curie Individual Fellowship No.~659316 (CYBERNETS) and in part by the U.S. National Science Foundation under Grants CNS-1702808 and ECCS-1647198.}
}
\maketitle

\begin{abstract}
	In this paper, the fundamental limits of decentralized simultaneous information and energy transmission in the $K$-user Gaussian multiple access channel (G-MAC), with an arbitrary $K \geqslant 2$ and one non-colocated energy harvester (EH), are fully characterized. 
		The objective of the transmitters is twofold. First, they aim to reliably communicate their message indices to the receiver; and second, to harvest energy at the EH at a rate not less than a minimum rate requirement $b$. 
The information rates $R_1,\dots,R_K$, in bits per channel use, are measured at the receiver and the energy rate $B$ is measured at an EH.
	Stability is considered in the sense of an $\eta$-Nash equilibrium ($\eta$-NE), with $\eta > 0$. 
	The main result is a full characterization of the $\eta$-NE information-energy region, i.e., the set of information-energy rate tuples $(R_1,\dots,R_K,B)$ that are achievable and stable in the G-MAC when: 
	$(a)$ all the transmitters autonomously and independently tune their own transmit configurations seeking to maximize their own information transmission rates $R_1,\dots, R_K$; and
	$(b)$ all the transmitters jointly guarantee an energy transmission rate $B$ at the EH, such that $B \geqslant b$. 
	Therefore, any rate tuple outside the $\eta$-NE region is not stable as there always exists at least one transmitter able to increase by at least $\eta$ bits per channel use its own information transmission rate by updating its own transmit configuration.
	\end{abstract}

\section{Introduction}
Recent years have witnessed a proliferation of battery-powered communication networks and devices. Within this context, when networks are deployed and batteries become inaccessible for either recharging or replacing, the network lifetime is often determined by the initial amount of energy stored in the batteries.
Therefore, when renewable energy sources such as light, wind, vibrations, etc., are not available, radio-frequency (RF) energy harvesting becomes an interesting alternative.
The main enabler of RF energy harvesting is the fact that energy and information can be simultaneously transmitted by radio-frequency waves, as first proposed by Tesla~\cite{Tesla1914-Patent,Wheeler-1943}. From this point of view, wireless networks can be designed to meet two objectives: $(a)$ information transmission to conventional receivers; and $(b)$ energy transmission to wireless energy harvesters (EHs). This two-objective paradigm for designing wireless networks is referred to as \emph{simultaneous information and energy transmission (SIET)} \cite{BelhadjAmor-ICT-2016}.

The fundamental limits on SIET are well-understood in point-to-point channels. The trade-off between the information transmission rate  and the energy transmission rate is characterized by the information-energy capacity function~\cite{VAR12}.
Consider for instance a binary symmetric channel (BSC) with cross-over probability $p$, i.e., $P(1|1)=P(0|0)=1-p$ and  $P(1|0)=P(0|1)=p$ and assume that the symbol `1' provides $1$ energy unit whereas the symbol `0' provides $0$ energy units. The information capacity of this channel is $1-H_2(p)$ bits per channel use, where $H_2(\cdot)$ denotes the binary entropy function. Note that the information capacity is achieved by equiprobable inputs, which induce an energy rate of $\frac12$ energy units per channel use at the EH.
The maximum energy rate is $1-p$, when `1' is always sent. 
\begin{figure}[t]
	\begin{center}
		\includegraphics[scale=0.7]{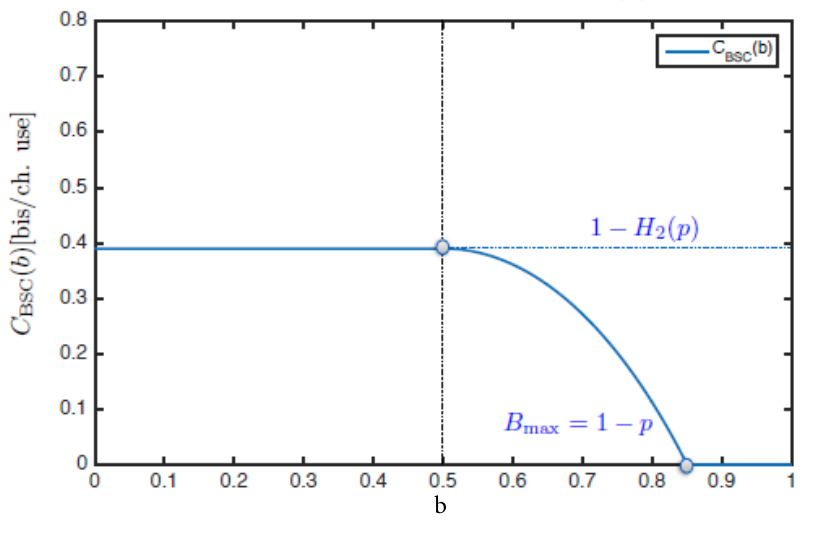}
		\caption{Information-Energy capacity function of $\text{BSC}(p)$ with $p=0.15$.\label{fig:BSC}}
	\end{center}
\end{figure}
Let $b$, with $0\leqslant b \leqslant 1-p$, denote the required minimum energy rate at the energy harvester. If $b>\frac12$, then equiprobable capacity-achieving inputs are not sufficient to achieve the minimum energy rate and 
the transmitter is forced to use the symbol `1' more frequently than the symbol `0', which induces an information rate loss.
In this case, the maximum information rate that can be achieved is $H_2(b)-H_2(p)$ and is strictly smaller than the information capacity. Note also that this maximum information rate is decreasing in $b$. Fig.~\ref{fig:BSC} depicts the information-energy capacity function for a BSC with crossover probability $p=0.15$. Note that if the two symbols carry the same energy per channel use, this trade-off is not observed.

In the context of multi-user channels, the information-energy fundamental limits are fully described by the \emph{information-energy capacity region}. That is, the set of all achievable information-energy rate tuples at which energy and information can be reliably transmitted.
The information-energy capacity region  in the discrete memoryless multi-access channel (MAC) and multi-hop networks was studied by Fouladgar \textit{et al.}~\cite{Fouladgar-CL-2012}. Recently, Belhadj Amor \textit{et al.}~\cite{BelhadjAmor-TIT-2017, BelhadjAmor-COMNET-2015} derived the exact information-energy capacity region of the Gaussian MAC (G-MAC) with an external EH in the cases with and without feedback.
Khalfet~\textit{et al.} ~\cite{Khalfet-Globecom,Khalfet-Allerton} derived the exact information-energy capacity region of the Gaussian interference channel (G-IC) with an external EH with and without feedback.
Analogously to the point-to-point case, these works show that there exist two energy regimes: one in which the energy rate constraint does not have any significant impact, and thus the set of achievable information rate tuples are those of the classical G-MAC or G-IC, respectively. Alternatively, in the other regime, increasing the information rates implies reducing the energy rate and \emph{vice-versa}. An object of central interest regarding the results in \cite{Fouladgar-CL-2012, BelhadjAmor-TIT-2017, BelhadjAmor-COMNET-2015,Khalfet-Globecom} and \cite{Khalfet-Allerton}, is that  the achievability of these information-energy rate tuples is subject to the existence of a central controller that decides an operating point and indicates to all network components the corresponding transmit-receive configuration that should be used. Unfortunately, this assumption does not hold in networks in which a central controller is not feasible. This is typically the case of \emph{decentralized} or \emph{ad hoc} networks such as sensor networks, body area networks, among others. In this type of multi-user channels, the transmitters and the receivers are assumed to be autonomous and capable of unilaterally choosing their own transmit-receive configurations, aiming to maximize their individual benefit, e.g., 
individual information rate, individual energy rate or a combination of both. Hence, from this perspective, the notion of information-energy capacity does not properly model  the fundamental limits of SIET in decentralized networks. 

To tackle this anarchical behavior observed in decentralized networks, a new condition on all achievable rate tuples is imposed: stability. There are many notions of stability, e.g., Nash equilibrium \cite{Nash-PNAS-1950}, Stackelberg equilibrium \cite{Stackelberg}, correlated equilibrium \cite{Aumann}, and satisfaction equilibrium \cite{Perlaza-JSSP2012}, among others. The remaining of this paper focuses exclusively on the $\eta$-Nash equilibrium ($\eta$-NE)~\cite{Nash-PNAS-1950}, with $\eta> 0$. A multi-user channel is stable in the sense of an $\eta$-NE if none of the transmitters or the receivers is able to increase its own individual benefit by more than $\eta$ units when unilaterally changing its own strategy. 
The set of all information-energy rate tuples that are achievable and stable is called the \emph{$\eta$-NE information-energy  region}.

\subsection{Previous Works}
Previous works have studied decentralized MACs using game-theoretic tools when the aim of each transmitter is limited to exclusively transmitting information. For instance,  
Lai and El Gamal \cite{Lai} proposed a framework to study the power allocation problem in fading decentralized MACs when the transmitters aim to maximize their own individual transmission rate. 
Gajic and Rimoldi \cite{GajicRimoldi} considered a similar scenario with time-invariant channels in which the transmitters have the choice of adopting any possible transmit configuration and determined the subregion of the information capacity region that is achievable at an NE.   
Belhadj Amor and Perlaza \cite{Netgcoop-2016} studied the $K$-user Gaussian MAC and characterized the fundamental limits of decentralized information transmission for two scenarios: a first game in which the transmitters autonomously and
independently tune their transmit configurations seeking to maximize their own
transmission rates, while the receiver adopts
a fixed receive configuration and stability is considered in the sense
of the $\eta$-NE, with $\eta> 0$; and  a second game involving the transmitters and the receiver, in which two categories of players play in a given order and stability is considered in the sense
of the $\eta$-sequential equilibrium~\cite{Breton-JOA-1988}, with with $\eta> 0$.
Varan and Yener~\cite{Varan} studied two-hop networks in which the source(s) is (are) incentivized to perform energy and signal cooperation to maximize  the amount of its (their) own data that is reliably delivered to the destination.

\subsection{Contributions}
This paper studies the fundamental limits  of \emph{decentralized SIET} in the two-user G-MAC when a minimum energy rate is required for successful decoding. 
More specifically, each transmitter chooses its own transmit configuration aiming to maximize its individual information rate to the receiver/information decoder while it guarantees an energy transmission rate higher than a given predefined threshold at a given EH. 
The receiver is assumed to adopt a fixed configuration that can be either single-user decoding (SUD), successive interference cancellation (SIC) or any time-sharing configuration of the previous decoding techniques. 
This paper provides a game formulation of this problem. The main contribution is the full characterization of the $\eta$-NE information-energy region of this game, with $\eta> 0$.

\subsection{Structure of the Paper}
The remainder of the paper is structured as follows. 
Section~\ref{sec:GaussiamMAC} describes the channel model and provides a game-theoretic formulation of decentralized SIET in the $K$-user G-MAC with a minimum energy rate constraint $b$. Section~\ref{SecNEregion} shows the main results of this paper and reports important observations. In Section~\ref{sec:proof_thm1} and Section~\ref{sec:proof_thm2}, the proofs are provided. Finally, Section~\ref{sec:conc} concludes the paper.

\section{$K$-User Gaussian MAC with Minimum Energy Rate $b$}
\label{sec:GaussiamMAC}
\begin{figure}[t]
	\centering{
		\begin{tikzpicture}[line cap=round,>=triangle 45,x=1.0cm,y=1.0cm,xscale=1.1,yscale=0.8,thick]
		\draw  (-0.5,3.1) rectangle (1.5,3.9);
		\draw  (-0.5,2.1) rectangle (1.5,2.9);
		\draw [->](1.5,3.5)--(2.35,3.5);
		\draw [->](1.5,2.5)--(2.35,2.5);
		\draw [->](2.65,2.5)--(3.4,0.6);
		\draw [->](2.6,3.5)--(3.35,3.5);
		\draw [->](2.6,3.5)--(3.5,0.65);
		\draw [->](1.5,0.5)--(2.36,0.5);
		\draw [->](2.6,0.5)--(3.35,0.5);
		\draw [->](2.6,0.5)--(3.5,3.35);
		\draw  (-0.5,0.1) rectangle (1.5,0.9);
		\draw[dotted] (4.4,0) rectangle (6.1,4);
		\draw (4.5,3.1) rectangle (6,3.9);
		\draw (4.5,0.1) rectangle (6,0.9);
		\draw [->](3.65,3.5)--(4.5,3.5);
		\draw [->](2.65,2.5)--(3.4,3.4);
		\draw [->](3.65,0.5)--(4.5,0.5);
		\draw [->](3.5,4.2)--(3.5,3.6);
		\draw [->](3.5,-0.25)--(3.5,0.35);
		\begin{small}
		\node at (0.5,3.5){Transmitter~1};
		\node[left] at (-0.5,3.5){$M_1$};
		\node at (0.5,2.5){Transmitter~2};
		\node[left] at (-0.5,2.5){$M_2$};
		\node[above right] at (2.5,3.5){$h_{11}$};
		\node at (3.5,1.5){$h_{21}$};
		\node at (2.5,1){$h_{1K}$};
		\node at (2.6,2){$h_{22}$};
		\node at (2.95,3.2){$h_{12}$};
		\node[above] at (2,3.5){$x_{1,t}$};
		\node[above] at (2,2.5){$x_{2,t}$};
		\node[below right] at (2.5,0.5){$h_{2K}$};
		\node[below] at (2,0.5){$x_{K,t}$};
		\node at (0.5,0.5){Transmitter~$K$};
		\node[left] at (-0.5,0.5){$M_K$};
		\node [right] at (3.5,4.22){$Z_t$};
		\node [right] at (3.5,-0.28){$Q_t$};
		\node[right] at (6,3.55){$(\hat{M}_1^{(n)},\dots ,\hat{M}_K^{(n)})$};
		\node at (5.25,3.5){Receiver};
		\node at (5.25,0.6){Energy};
		\node at (5.25,0.3){Harvester};
		\node at (6.7,0.5){Energy};
		\node[below] at (3.9,3.5){$Y_{1,t}$};
		\node[above] at (3.9,0.4){$Y_{2,t}$};
		\end{small}
		\begin{large}
		\node at (2.48,3.5){$\otimes$};
		\node at (0.5,1.6){$\vdots$};
		\node at (1.8,1.6){$\vdots$};
		\node at (2.48,2.5){$\otimes$};
		\node at (2.48,0.5){$\otimes$};
		\node at (3.5,3.5){$\oplus$};
		\node at (3.5,3.5){$\oplus$};
		\node at (3.5,0.5){$\oplus$};
		\end{large}
		\end{tikzpicture}}
	\caption{$K$-user memoryless Gaussian MAC with  energy harvester.
	}
	\label{FigMAC-FB}
\end{figure}
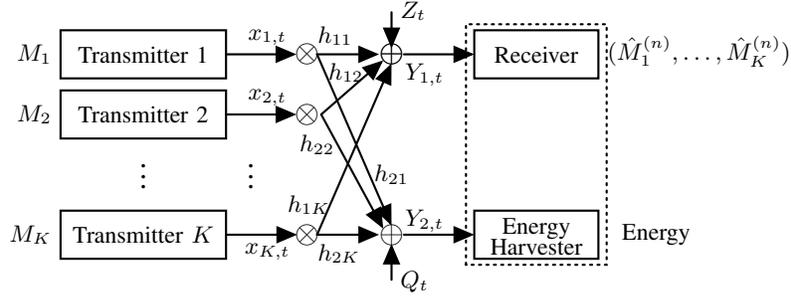
\subsection{Channel Model}
Consider the $K$-user memoryless Gaussian multiple access channel (G-MAC) with an energy harvester (EH), with an arbitrary number of users $K\geqslant 2$, as shown in Fig.~\ref{FigMAC-FB}.
Let $n \in \mathds{N}$ be the blocklength. At each channel use $t\in \{1,\dots, n\}$ and for any $i\in \{1,\dots, K\}$, let $X_{i,t}$  denote the real symbol sent by transmitter~$i$. 
The receiver observes the real channel output 
\begin{equation}
\label{EqY}
Y_{1,t}=\sum_{i=1}^K h_{1i} X_{i,t} + Z_t,
\end{equation}
and the EH observes 
\begin{equation}
\label{EqS}
Y_{2,t} =\sum_{i=1}^K h_{2i} X_{i,t} + Q_t,
\end{equation}
where $h_{1i}$ and $h_{2i}$ are the corresponding constant non-negative channel coefficients from transmitter $i$ to the receiver and EH, respectively. 

The noise terms $Z_{t}$ and $Q_{t}$ are realizations of two identically distributed
zero-mean real Gaussian random variables with variances $\sigma_1^2$ and $\sigma_2^2$, respectively. In the following, there is no particular assumption on the joint distribution of $Q_t$ and $Z_t$.

The symbols $\{X_{i,t}\}_{t=1}^n$ satisfy an expected average \emph{input power constraint}
\begin{IEEEeqnarray}{C}
	\label{EqPowerConstraint}
	P_i = \frac1n \sum_{t=1}^n \E{X_{i,t}^2} \leqslant P_{i,\max},
\end{IEEEeqnarray}
where $P_i$ and $P_{i,\max}$ denote respectively the average transmit power and the maximum average power of transmitter~$i$ in energy units per channel use, with $i\in\{1,\dots,K\}$.

The channel coefficients satisfy the following $\set{L}_2$-norm condition:
\begin{equation} 
\forall j\in \{1,2\},\quad \|\vect{h}_j\|^2 \leqslant 1,
\end{equation} 
with $\vect{h}_j\eqdef \trans{(h_{j1},\dots,h_{jK})}$, in order to satisfy the energy conservation principle.

The signal to noise ratios (SNRs): $\SNR_{ji}$, with $\forall (j,i) \in \{1,2\}\times \{1,\dots,K\} $ are defined as follows
\begin{equation}
\SNR_{ji}\eqdef |h_{ji}|^2 \frac{P_{i,\max}}{\sigma_j^2}.
\end{equation}
%
%
%
Within this context, two main tasks are to be simultaneously accomplished: information transmission and energy transmission.
\subsection{Information Transmission}
The goal of the communication is to  convey the independent message $M_i$,  uniformly distributed over the set $\set{M}_i\triangleq\{1,\dots, \lfloor 2^{n R_i}\rfloor\}$, from transmitter~$i$, with $i\in \{1,\dots,K\}$, to the common receiver at the information rate $R_i$.  The message indices $(M_1,\dots,M_K)$ are independent of the noise terms $Z_{1}, \ldots, Z_{n}$ and  $Q_{1}, \ldots, Q_{n}$. 

At each time $t$,  the $t$-th symbol of transmitter~$i$, for ${i\in\{1,\dots,K\}}$, depends solely on its message index $M_i$ and a randomly generated index $\Omega \in \{1,\dots, \lfloor 2^{n R_{r}}\rfloor\}$, with $R_r \geqslant 0$, that is independent of $M_1,\dots,M_K$ and assumed to be known by all transmitters and by the receiver, i.e., 
\begin{IEEEeqnarray}{rCL}
	X_{i,t}&=&f_{i,t}^{(n)} (M_i, \Omega),\quad t\in\{1,\dots,n\},
\end{IEEEeqnarray}
for some encoding functions 
$f_{i,t}^{(n)} \colon  \set{M}_i  \times \mathds{N} \to \mathbb{R}$.
%
The receiver produces an estimate $(\hat{M}_1^{(n)},\dots,\hat{M}_K^{(n)})=\Phi^{(n)}(Y_{1,1},\dots,Y_{1,n},\Omega)$ of the message tuple $(M_1,\dots ,M_K)$ via a decoding function 
$\Phi^{(n)}\colon \mathbb{R}^{n}\times \mathbb{N}\to \set{M}_1\times\dots \times\set{M}_K$, 
and the average probability of error is given by
\begin{IEEEeqnarray}{rCL} 
	P_{\error}^{(n)} (R_1, \dots,R_K) &\eqdef&\Pr\left[ (\hat{M}_1^{(n)},\dots, \hat{M}_K^{(n)}) \neq (M_1,\dots,M_K)\right].
\end{IEEEeqnarray}

\subsection{Energy Transmission}
Let $b_{\textnormal{coop}}(K)\geqslant0$ denote the maximum energy rate that can be achieved at the input of the EH given the input power constraints. It is given by
\begin{equation}
\label{MaxFeasibleEnergy}
b_{\textnormal{coop}}(K)\eqdef 1 + \left(\sum_{i=1}^K\sqrt{\SNR_{2i}}\right)^2.
\end{equation}
This rate can be achieved when the transmitters use all their power budgets to send fully correlated channel inputs.

Let also $b_{\textnormal{ind}}(K)\geqslant 0$  denote the maximum energy rate that can be achieved at the input of the EH given the input power constraints when the channel inputs are independent. 
It is given by
\begin{equation}
\label{MaxIndEnergy}
b_{\ind}(K)\eqdef 1 + \sum_{i=1}^K\SNR_{2i}.
\end{equation}

Let $b \geqslant 0$ denote the minimum energy rate that must be guaranteed at the input of the EH in the G-MAC. 
This rate $b$ must satisfy
\begin{equation}
0\leqslant b \leqslant b_{\coop}(K),\label{EqFeasibleb}
\end{equation}
for the problem to be feasible.

The empirical energy transmission rate induced by the sequence $(Y_{2,1},\dots,Y_{2,n})$ at the input of the EH is 
\begin{IEEEeqnarray}{ccl}\label{EqEH}
	B^{(n)}&\eqdef& \frac{1}{n}\sum_{t = 1}^{n} \frac{Y_{2,t}^2}{\sigma_2^2},
\end{IEEEeqnarray} 
given the normalization over the noise power.

The goal of the energy transmission is to guarantee that the empirical energy rate $B^{(n)}$ is not less than an operational energy rate $B$ that must satisfy 
\begin{equation}
\label{EqB}
b \leqslant B \leqslant b_{\coop}(K).
\end{equation}

Hence, the probability of energy outage is defined as follows:
\begin{IEEEeqnarray}{rCl}
	\label{EqEnergyOutage}
	P_\outage^{(n)} (B) &=&\Pr\left[B^{(n)}<B-\epsilon\right],
\end{IEEEeqnarray}
for some $\epsilon>0$ arbitrarily small.

To ease notation, the acronym \GMAC is used to refer to the $K$-user G-MAC  with an EH depicted in Fig.~\ref{FigMAC-FB} with fixed SNRs: $\SNR_{ji}$, for all $(j,i)\in \{1,2\}\times \{1,\dots,K\}$ and minimum energy rate requirement $b$.

Without loss of generality, it can be assumed that $\sigma_1^2=\sigma_2^2=1$.

\subsection{Centralized Simultaneous  Information and Energy Transmission}

The G-MAC depicted in Fig.~\ref{FigMAC-FB} is said to operate at the information-energy rate tuple $(R_1,\dots ,R_K, B) \in \mathds{R}_{+}^{K}\times [b,\infty)$ when all the transmitters and the receiver use a transmit-receive configuration such that:
$(i)$ information transmission occurs at rates $R_1, \dots, R_K$ with probability of error arbitrarily close to zero; and 
$(ii)$ energy transmission occurs at a rate not smaller than $B$ with energy-outage probability arbitrarily close zero.
Under these conditions, the information-energy rate tuple $(R_1,\dots ,R_K, B)$ is said to be achievable in the\!\GMAC\!. 
\begin{definition}[Achievable Rates]\label{DefAchievableTriples}
	In the\!\GMAC\!, the information-energy rate tuple  $(R_1,\dots,R_K, B) \in \mathds{R}_{+}^{K}\times [b,\infty)$ is achievable if there exists a sequence of encoding and decoding functions  $\big\{\{f_{1,t}^{(n)}\}_{t=1}^n,\dots,\{f_{K,t}^{(n)}\}_{t=1}^n,\Phi^{(n)}\big\}_{n=1}^\infty$ such that both the average error probability and the energy-outage probability tend to zero as the blocklength $n$ tends to infinity. That is,
	\begin{IEEEeqnarray}{lcl}
		\label{EqProbError}
		\limsup_{n \rightarrow \infty}\;  P_{\error}^{(n)}(R_1,\dots,R_K)  & = & 0, \mbox{ and }\\
		\label{EqProbPower}
		\limsup_{n \rightarrow \infty}\;  P_\outage^{(n)} (B)& = & 0.
	\end{IEEEeqnarray}
\end{definition}
Note that the minimum energy rate constraint $b$ requires in particular that
\begin{equation}
\label{EqConditionb}
\limsup_{n \rightarrow \infty}\;  P_\outage^{(n)} (b) = 0.
\end{equation}
Often, increasing the energy transmission rate implies decreasing the information transmission rates and \emph{vice-versa}.
An important notion to characterize the fundamental limits on this information-energy trade-off is the \emph{information-energy capacity region} defined as follows: 
\begin{definition}[Information-Energy Capacity Region]\label{DefCER}
The information-energy capacity region $\set{E}(K,b)$ of the\GMAC is the closure of all achievable information-energy rate tuples $(R_1, \dots,R_K, B)$.
\end{definition}
The information-energy capacity region $\set{E} (K,b)$ is described by the following theorem.
\begin{theorem}[Information-Energy Capacity Region]
	\label{TheoremCentralizedRegion}
	The information-energy capacity region $\set{E}(K,b)$ of the \GMAC is the set of all information-energy rate tuples $(R_1, \dots,R_K, B)$ that satisfy
	\begin{subequations}
		\label{eq:regprop1}	
		\begin{IEEEeqnarray}{ccccl}
0 &\leqslant & \sum_{j\in \set{U}} R_j & \leqslant & \frac{1}{2} \log_2 \big( 1 + \sum_{j\in \set{U}}  \beta_j\;\SNR_{1j} \big), \quad \forall \set{U} \subseteq \{1,\dots,K\},\label{eq:thm1_c12}\\
			b&\leqslant & B  & \leqslant &  1 + \sum_{j=1}^K \beta_j\SNR_{2j} + \left(  \sum_{j=1}^K   \sqrt{(1-\beta_j)\SNR_{2j}} \right)^2,\label{eq:thm1_e}
		\end{IEEEeqnarray}
	\end{subequations}
	with $(\beta_1,\dots,\beta_K) \in \left[ 0, 1 \right]^K $. 
\end{theorem}
\begin{IEEEproof}
	The proof of Theorem \ref{TheoremCentralizedRegion} follows immediately from \cite[Proposition~1]{BelhadjAmor-TIT-2017} and \cite[Theorem~2]{BelhadjAmor-TIT-2017} when generalized to $K$ users.
\end{IEEEproof}

\textbf{Comments and Observations:} In the constraints~\eqref{eq:regprop1}, when the parameters $\beta_1,\dots,\beta_K$ satisfy $\beta_1=\dots=\beta_K=1$, the corresponding region is characterized by all information-energy rate tuples $(R_1, \dots,R_K, B)$ that satisfy
\begin{subequations}
	\label{eq:regbeta1}	
	\begin{IEEEeqnarray}{ccccl}
		0 &\leqslant & \sum_{j\in \set{U}} R_j & \leqslant & \frac{1}{2} \log_2 \big( 1 + \sum_{j\in \set{U}}  \SNR_{1j} \big), \quad \forall \set{U} \subseteq \{1,\dots,K\},\label{eq:18a}\\
		b&\leqslant & B  & \leqslant &  b_\ind (K)\label{eq:18b}.
	\end{IEEEeqnarray}
\end{subequations}
That is, the information rate constraints in \eqref{eq:18a} describe the capacity region of the $K$-user G-MAC and the upper bound on the energy rate constraint in \eqref{eq:18b} corresponds to the maximum energy rate that can be achieved using independent channel inputs. 

On the other hand, when the parameters $\beta_1,\dots,\beta_K$ are such that $\beta_1=\dots=\beta_K=0$, the corresponding region is characterized by all information-energy rate tuples $(R_1, \dots,R_K, B)$ that satisfy
\begin{subequations}
	\label{eq:regbeta0}	
\begin{IEEEeqnarray}{l}
R_1=\dots=R_K=0,\label{eq:19a}\\		
b\leqslant  B   \leqslant  b_{\coop}(K).\label{eq:19b}
\end{IEEEeqnarray}
\end{subequations}
That is, the information rate constraints in \eqref{eq:19a} do not allow for any information transmission and the upper bound on the energy rate $B$ in \eqref{eq:19b} equals the maximum feasible energy rate.

Hence, from this constructive viewpoint, the coefficients $\beta_1,\dots,\beta_K$ in~\eqref{eq:regprop1} allow the transmitters to trade off between information and energy rates. These parameters might be interpreted as the fractions of power that the transmitters allocate for information transmission. At each transmitter $i$, the remaining fraction of power $(1-\beta_i)$ is allocated  for  exclusively transmitting energy to the EH. More specifically, to achieve any information-energy rate tuple in this region, at each time~$t$, transmitter~$i$'s  channel input can be written as:
\begin{IEEEeqnarray}{rCl}
	\label{EqCodingScheme}
	X_{i,t}=\sqrt{(1-\beta_i) P_i} W_t+U_{i,t},\quad i \in \{1,\dots,K\},
\end{IEEEeqnarray}
for some independent zero-mean Gaussian information-carrying symbols $U_{1,t},\dots, U_{K,t}$ with variances $\beta_1 P_1,\dots,\beta_K P_K$, respectively, and independent thereof $W_t$ are  zero-mean unit-variance Gaussian energy-carrying symbols known non-causally to all terminals. The codebook and the encoding-decoding schemes for the information-carrying signals can be those described in \cite{COVER75} or \cite{WYNER74}.

Note that the information-carrying signals carry both energy and information. These signals are useful to both the EH and the information decoder,  whereas the other signals are energy-carrying and are useful only to the EH. These energy-carrying signals carry only common randomness that allows the creation of correlated signals to increase the energy rate. The common randomness is known to the information decoder and does not produce any interference to the information-carrying signals as its effect can be suppressed using classical successive interference cancellation.

		An inherent assumption here is the existence of a central controller that determines an operating point and imposes the transmit or receive configuration to be adopted by each network component. From this global or centralized perspective all points inside the information-energy capacity region are possible operating points.
		However, in a decentralized network, each network component is an autonomous decision maker that aims to maximize its own individual reward by appropriately choosing a particular transmit or receive configuration. From this perspective, only the information-energy rate tuples that are \emph{stable} can be possible asymptotic operating points. 
	
	The following subsection describes decentralized SIET in the G-MAC$(K,b)$.

\subsection{Decentralized Simultaneous Information and Energy Transmission}

In a decentralized G-MAC$(K,b)$, the aim of transmitter $i$, for all $i\in \{1,\dots, K\}$, is to autonomously choose its transmit configuration $s_i$ in order to maximize its information rate $R_i$, while guaranteeing a minimum energy rate $b$ at the EH. In particular, the transmit configuration $s_i$ can be described in terms of the information rates $R_i$, the block-length $n$, the channel input alphabet $\mathcal{X}_i$, the encoding functions $\{f_{i,t}^{(n)}\}_{t=1}^n$, the common randomness, the power dedicated to information and energy transmission, etc. 
The receiver is assumed to adopt a fixed decoding strategy that is known in advance to all transmitters.  

Let $\mathcal{P}_K$ denote the set of all permutations (all possible decoding orders) over the set $\{1,\dots,K\}$.  For any $\pi\in \mathcal{P}_K$, the considered decoding order $\pi(1),\pi(2),\dots,\pi(K)$ is such that user $\pi(1)$ is decoded first, user $\pi(2)$ is decoded second, etc.

Note that if the aim of each transmitter, say transmitter $i$, is to maximize its own individual information rate $R_i$ subject to the minimum energy rate $b$ at the EH, it is clear from \eqref{eq:regprop1} that one option should be using a power-split in which the component dedicated to the transmission of information $\beta_i$ is as high as possible. However, its power-split $\beta_i$ must also be chosen such that the energy-outage probability \eqref{EqConditionb} can be made arbitrarily close to zero. 
This reveals that the choice of the transmit configuration of  each transmitter depends on the choice of the other transmitters as they must guarantee the minimum energy constraint required at the EH. At the same time, depending on the decoding scheme at the receiver, the information-carrying signal of one transmitter is interference to the others.
This reasoning implies that the rate achieved by transmitter $i$ depends on all transmit-configurations $s_1,\dots,s_K$ as well as the configuration  of the receiver, even if it is assumed to be fixed. This justifies the analysis of this scenario using tools from game theory. 

\subsection{Game Formulation}\label{SecGameFormulation}
The competitive interaction of the transmitters in the decentralized\GMAC can be modeled by the following game in normal form:
\begin{equation}\label{EqGame}
\GameNF,
\end{equation}
where $b$ is a parameter of the game that represents the minimum energy-rate that must be guaranteed at the EH (see \eqref{EqConditionb}).
The set $\mathcal{K} = \lbrace 1,\dots,K \rbrace$ is the set of players, i.e., transmitters $1$ to $K$. The sets $\mathcal{A}_1,\dots,\mathcal{A}_K$ are the sets of actions of players  $1$ to $K$, respectively. An action of a player $i \in \mathcal{K}$, which is denoted by $s_i \in \mathcal{A}_i$, is basically its transmit configuration as described above. 
The utility function of transmitter~$i$, for $i\in \set{K}$, is $u_i: \mathcal{A}_1\times  \dots\times \mathcal{A}_K \rightarrow \mathds{R}$ and it is defined  as its own information rate,
\begin{equation}
\label{EqUtilityTX}
\small
u_i(s_1,\dots,s_K) = \left\lbrace
\begin{array}{lcl}
R_i(s_1,\dots,s_K)=\frac{\log_2|\set{M}_i|}{n}, 	& \mbox{if }  	& P_{\error}^{(n)}(R_1,\dots,R_K) < \epsilon \text{ and } P_\outage^{(n)}(b)< \delta\\
-1, 			& 			&\mbox{otherwise,}
\end{array}\right.
\end{equation} 
where $\epsilon > 0$ and $\delta > 0$ are arbitrarily small numbers and $R_i(s_1,\dots,s_K)$ denotes an information transmission rate achievable (Def.~\ref{DefAchievableTriples}) with the configurations $s_1,\dots,s_K$. 
Note that the utility is -1 when either the error probability or the energy outage probability is not arbitrarily small. This is meant to favor the action profiles in which there is no information transmission (information rate and error probability are zero) but there is energy transmission (probability of energy outage can be made arbitrarily close to zero), over the actions in which the information rate is zero but the energy constraint is not satisfied.

\begin{remark}	
	A player wants to maximize its individual information rate while guaranteeing a global energy rate (i.e., common to all players).  Hence, from player $i$'s standpoint, with $i\in\set{K}$, what matters is the error probability on decoding message $M_i$, i.e, its individual error probability. Although the game formulation presented in this paper relies on the joint probability of error, one can argue that it can equivalently be written in terms of individual error probabilities because 
	\begin{equation}\max_{i\in\set{K}} \Pr\left[\hat{M}_i^{(n)}\neq M_i \right]\leqslant \Pr\left[ (\hat{M}_1^{(n)},\dots, \hat{M}_K^{(n)}) \neq (M_1,\dots,M_K)\right]\leqslant K\cdot \max_{i\in\set{K}} \Pr\left[\hat{M}_i^{(n)}\neq M_i \right].	
	\end{equation}
\end{remark}
Often, the information rate $R_i(s_1,\dots,s_K)$ is  written as $R_i$ for simplicity. However, every non-negative achievable information rate is associated with a particular tuple of transmit configurations $(s_1,\dots,s_K)$ that achieves it. It is worth noting that there might exist several tuples of transmit configurations that achieve the same rate tuple $(R_1, \dots,R_K, B)$ and distinction between the different transmit configurations tuples is made only when needed. 
Note also that the transmitters could use different blocklengths $n_i$, with $i\in\set{K}$, as part of the set of actions of each player. However, the decoding is performed only after all blocks are received. As a consequence, without loss of generality the blocklength can be considered to be $n=\ds\max_{i\in \set{K}}n_i$ for all the players.

A class of transmit configuration tuples $\boldsymbol{s}^* = (s_1^*,\dots, s_K^*) \in \mathcal{A}_1 \times\dots\times \mathcal{A}_K$ that are particularly important in the analysis of this game are referred to as $\eta$-Nash equilibria ($\eta$-NE).

\subsection{$\eta$-Nash Equilibria}
A transmit-receive configuration $\boldsymbol{s}^* = (s_1^*, \dots,s_K^*) \in \mathcal{A}_1 \times\dots\times  \mathcal{A}_K$ that is an $\eta$-NE  satisfies the following definition:
\begin{definition}[$\eta$-NE \cite{Nisan-Book-2007}] \label{DefEtaNE} \emph{
		In the game $\gameNF$, an action profile  $(s_1^*, \dots,s_K^*)$ is an $\eta$-NE if for all $i\in \set{K}$ and for all $s_i \in \mathcal{A}_i$, it holds that
		\begin{IEEEeqnarray}{lcl}\label{EqNashEquilibrium} 
			u_i(s_1^*,\dots,s_{i-1}^*,s_i,s_{i+1}^*, \dots,s_K^*)  &  \leqslant & u_i(s_1^*,\dots,s_{i-1}^*,s_i^*,s_{i+1}^*, \dots,s_K^*) + \eta.
		\end{IEEEeqnarray}
	}
\end{definition}

From Def. \ref{DefEtaNE}, it becomes clear that if $(s_1^*,\dots ,s_K^*)$ is an $\eta$-NE, then none of the transmitters can increase its own information transmission rate  by  more than $\eta$ bits per channel use by changing its own transmit configuration while keeping the average error probability and the energy outage probability arbitrarily close to zero. Thus, at a given $\eta$-NE, every player achieves a utility that is $\eta$-close to its maximum achievable information rate given the energy rate requirement $b$, the receive configuration of the information decoder, and the transmit configuration of the other players. 
Note that if $\eta$ is arbitrarily close to zero, then the definition of NE  is obtained~\cite{Nash-PNAS-1950}. 

The following investigates the set of information and energy rate tuples that can be achieved at an $\eta$-NE. This set of rate tuples is known as the $\eta$-NE information-energy region.

\begin{definition}[$\eta$-NE Region] \label{DefNERegion} \emph{
		Let $\eta> 0$ be fixed. An achievable information-energy rate tuple $(R_1,\dots,R_K,B)\in \set{E}(K,b)$ is said to be in the $\eta$-NE region of the game $\gameNF$ if there exists an action profile $(s_1^*,\dots ,s_K^*) \in \mathcal{A}_1 \times\dots\times \mathcal{A}_K$  that is  an  $\eta$-NE and the following holds:
		\begin{IEEEeqnarray}{ccl}
		u_i (s_1^* , \dots ,s_K^*)  &=& R_i, \quad \forall i \in \{1,\dots,K\}.
		\end{IEEEeqnarray}
	}
\end{definition}

In the following section, the $\eta$-NE region of the game $\gameNF$, with an $\eta > 0$,  is characterized for several decoding strategies adopted by the information decoder under a given feasible minimum energy rate constraint $b$ satisfying \eqref{EqFeasibleb}.

\section{Main Results}\label{SecNEregion}
This section describes the $\eta$-NE region of the game $\gameNF$ under a fixed decoding strategy. Three cases are examined: single user decoding (SUD), successive interference cancellation (SIC), and time-sharing between the previous decoding strategies.
\subsection{$\eta$-NE Region With Single User Decoding (SUD)}
The $\eta$-NE region of the game $\gameNF$ when the receiver uses single-user decoding (SUD), denoted by $\set{N}_{\mathrm{SUD}}(K,b)$, is described by the following theorem. 
\begin{theorem}[$\eta$-NE Region of the Game $\gameNF$ with SUD]\label{thm1}
	Let $b\in [0, b_\coop(K)]$ and $\eta > 0$ be fixed. Then, the set  $\set{N}_{\mathrm{SUD}}(K,b)$ of $\eta$-NE of the game $\gameNF$ contains all  information-energy rate tuples $(R_1,\dots,R_K,B)$ that satisfy
	\begin{subequations}	\label{cstThm1}
		\begin{align}
			 R_i &= \frac12 \log_2\left(1+\frac{\beta_i\SNR_{1i}}{1+\ds\sum_{j=1;j\neq i}^K \beta_j\SNR_{1j}}\right),\quad \forall i \in \{1,\dots,K\},\\
			b&\leqslant	B      \leqslant   1 + \sum_{j=1}^K \beta_j\SNR_{2j} + \left(  \sum_{j=1}^K   \sqrt{(1-\beta_j)\SNR_{2j}} \right)^2,
		\end{align}
	\end{subequations}
	where $\beta_1,\dots,\beta_K$ satisfy the following conditions:
	\begin{IEEEeqnarray}{l}
\beta_1=\dots=\beta_K=1\quad\text{when}\quad b\in \left[0,b_\ind(K)\right]; \text{ and}\\
	\label{eq:27}
	1 + \sum_{j=1}^K \beta_j\SNR_{2j} + \left(  \sum_{j=1}^K   \sqrt{(1-\beta_j)\SNR_{2j}} \right)^2=b,\quad\text{when}\quad b\in \left(b_{\ind}(K),b_{\coop}(K ) \right].
	\end{IEEEeqnarray}
 
\end{theorem}
\begin{IEEEproof}	
The proof of Theorem~\ref{thm1} is provided in Section~\ref{sec:proof_thm1}.\end{IEEEproof}
That is, when $b\in [0,b_\ind(K)]$, $\set{N}_\SUD(K,b)$ contains all information-energy rate tuples $(R_1,\dots,R_K,B)$ such that
\begin{subequations}
\begin{align}
R_i&=\frac12 \log_2\left(1+\frac{\SNR_{1i}}{1+\ds\sum_{j=1;j\neq i}^K \SNR_{1j}}\right),\quad \forall i \in \{1,\dots,K\},\\
b&\leqslant B \leqslant b_\ind (K).
\end{align}
\end{subequations}
Thus, any projection of $\set{N}_\SUD(K,b)$ over a plane $B=b_1$, with $b\leqslant b_1 \leqslant b_\ind (K)$, reduces to a unique information rate point (See Fig.~\ref{Fig2} in the case of two users). 

When $b\in (b_\ind(K),b_\coop(K) ]$,
there are infinitely many  tuples $(\beta_1,\dots,\beta_K)$ that satisfy \eqref{eq:27}. For a given choice of $(\beta_1,\dots,\beta_K)$, the constraints in \eqref{cstThm1} reduce to
\begin{subequations}
\begin{IEEEeqnarray}{cCl}
	R_i&=&\frac12 \log_2\left(1+\frac{\beta_i\SNR_{1i}}{1+\ds\sum_{j=1;j\neq i}^K\beta_j\SNR_{1j}}\right),\quad \forall i \in \{1,\dots,K\},\\
	B&=&b. 
	\end{IEEEeqnarray}
\end{subequations}
That is, at any $\eta$-NE, the energy rate must be satisfied with equality in order to maximize the information rates (See Fig.~\ref{Fig3} in the case of two users).
\subsection{$\eta$-NE Region with Successive Interference Cancellation (SIC)}

The $\eta$-NE region of the game $\gameNF$ when the information decoder uses $\SIC(\pi)$, with a fixed decoding order $\pi\in \set{P}_K$, denoted by $\set{N}_{\SIC(\pi)}(K,b)$,  is described by the following theorem. 
\begin{theorem}[$\eta$-NE Region of the Game $\gameNF$ with $\SIC(\pi)$]\label{thm2}
	Let $b\in [0, b_\coop(K)]$ and ${\eta > 0}$ arbitrarily small. Then, the set $\set{N}_{\SIC(\pi)}(K,b)$ contains all information-energy rate tuples $(R_1,\dots,R_K,B)$ satisfying:
	\begin{subequations}	\label{cstThm2}
		\begin{align}
			R_{\pi(i)} &= \frac12 \log_2\left(1+\frac{\beta_{\pi(i)}\SNR_{1\pi(i)}}{1+\sum_{j=i+1}^K \beta_{\pi(j)}\SNR_{1\pi(j)}}\right), \quad \forall i \in \{1,\dots,K\},\\
			b &\leqslant  B   \leqslant   1 + \sum_{j=1}^K \beta_j\SNR_{2j} + \left(  \sum_{j=1}^K   \sqrt{(1-\beta_j)\SNR_{2j}} \right)^2,
		\end{align}
	\end{subequations}
	where $\beta_1,\dots,\beta_K$ satisfy the following conditions:
\begin{align}
	&\beta_1=\dots=\beta_K=1\quad\text{when}\quad b\in [0,b_\ind(K)]; \text{ and}\\
	\label{eq:condthm2}
	& 1 + \sum_{j=1}^K \beta_j\SNR_{2j} + \left(  \sum_{j=1}^K   \sqrt{(1-\beta_j)\SNR_{2j}} \right)^2=b,\quad\text{when}\quad b\in (b_{\ind}(K),b_{\coop}(K ) ].
\end{align}
\end{theorem}
\begin{IEEEproof}The proof of  Theorem~\ref{thm2} is provided in Section~\ref{sec:proof_thm2}.\end{IEEEproof}
The observations in the previous case continue to hold here. The only difference is that the information rate constraint of transmitter $\pi(i)$ is not affected by the signals of transmitters $\pi(1)$ to $\pi(i-1)$ that were already decoded. Interference cancellation allows in particular the achievability of information sum-rate optimal points at an $\eta$-NE (points on the boundary of the information-energy  capacity region) as shown in Fig.~\ref{Fig2} and Fig.~\ref{Fig3} in the case of two users. 
\subsection{$\eta$-NE Region With Time-Sharing (TS)}
Let $\set{N}(K,b)$ denote the $\eta$-NE region of the game $\gameNF$ when the receiver uses any time-sharing  between the previous decoding techniques. This region is described by the following theorem. 

\begin{theorem}[$\eta$-NE Region of the Game $\gameNF$]\label{thm3}
	Let $b\in [0, b_\coop(K)]$ and ${\eta > 0}$ be fixed. Then, the set $\set{N}(K,b)$ is defined as:
	\begin{equation}
	\set{N}(K,b)=\textnormal{Convex hull}\left(\set{N}_{\mathrm{SUD}}(K,b) \cup \left( \bigcup_{\pi\in \set{P}_K}\set{N}_{\SIC(\pi)}(K,b)\right)\right).	
	\end{equation}
\end{theorem}
That is, if the receiver performs any time-sharing combination between any of the considered decoding strategies, then the transmitters can use the same time-sharing combination between their corresponding $\eta$-NE strategies to  achieve any point inside $\set{N}(K,b)$. 
\begin{IEEEproof}The proof is based on Theorem~\ref{thm1}, Theorem~\ref{thm2}, and a time-sharing argument. The details are omitted.\end{IEEEproof}

\subsection{Observations}

Three main observations arising from Theorem~\ref{thm1}, Theorem~\ref{thm2}, and Theorem~\ref{thm3} are described in the sequel.   
\subsubsection{Existence of an $\eta$-NE}
The first observation is that the existence of an $\eta$-NE, with $\eta>0$ fixed, is always guaranteed as long as the SIET problem is feasible, i.e., as long as $b \leqslant b_\coop(K)$. This statement follows immediately from the fact that $\set{N}_\SUD(K,b) \neq \emptyset$, $\set{N}_{\SIC(\pi)}(K,b) \neq \emptyset$, for any $\pi \in \set{P}_K$, and thus $\set{N}(K,b) \neq \emptyset$, which ensures the existence of at least one action profile $(s_1^*,\dots,s_K^*)$ that is an $\eta$-NE. Interestingly, when $b > b_\coop(K)$, the energy transmission cannot be performed reliably, and thus the information-energy capacity region is empty and so is the $\eta$-NE region.
In this particular case, the problem is not well-posed since such an energy rate is outside the information-energy capacity region.

\textbf{Remark:} Note that for any given $b \geqslant 0$, the sets $\set{N}_{\mathrm{SUD}}(K,b)$, $\set{N}_{\SIC(\pi)}(K,b)$, for $\pi\in \set{P}_K$, and $\set{N}(K,b)$ include only the information-energy tuples $(R_1,\dots,R_K,B)$ that satisfy $B \geqslant b$. That is, the $\eta$-NE at which the energy constraint can be satisfied. However, this suggests that there might exist other $\eta$-NE that are not in these sets at which either one of the conditions, \eqref{EqProbError} or \eqref{EqProbPower}, is not met. Consider for instance a case in which $b \geqslant 1 + \max_{i\in \set{K}} \SNR_{2i}$ and all the transmitters decide to use the strategies $s_1$ to $s_K$ at which none of the transmitters actually transmits, e.g., standby mode. Hence, none of the transmitters can unilaterally deviate and achieve a utility other than $u_{i}(s_1,\dots,s_K) = -1$, for all $i\in \{1,\dots,K\}$, which translates into an all-zero information-energy tuple $(0,\dots,0)$ which is also an $\eta$-NE but is not in any of the sets $\set{N}_{\SUD}(K,b)$ or $\set{N}_{\SIC(\pi)}(K,b)$, fo any $\pi\in \set{P}_K$, as the energy constraint cannot be satisfied (Def.~\ref{DefNERegion}). More specifically, $(0,\dots,0)\notin \set{E}(K,b)$ for all $b>0$.

\subsubsection{Cardinality of the set of $\eta$-NE}
The unicity of a given $\eta$-NE of the game $\gameNF$ is not ensured even in the case in which the cardinality of the  $\eta$-NE information-energy region is one. Consider the case in which $\eta$ is arbitrarily close to zero and $b =b_\coop(K)$. In this case, $\set{N}(K,b)= \lbrace (0,\dots,0,b_\coop(K)) \rbrace$ and using, for instance, all the power budget to send common randomness is an $\eta$-NE action profile, for any $\eta> 0$. However, there is an infinite number of possible common random sequences that can be adopted by the transmitters independently of the action taken by the receiver as in this case $R_1 =\dots= R_K = 0$. The cardinality of the set of $\eta$-NE is an acceptable lower-bound on the number of equilibria. This suggests that if the cardinality of the  $\eta$-NE set is  infinity, the number of $\eta$-NE is also infinity as every information-energy rate tuple in $\set{N}(K,b)$ is associated with at least one achievability scheme that is an $\eta$-NE (Def.~\ref{DefNERegion}). 

\subsubsection{Optimality of the $\eta$-NE}
The most interesting observation regarding Theorem \ref{thm1}, Theorem~\ref{thm2}, and Theorem~\ref{thm3} is that some of the sum-rate optimal tuples $(R_1,\dots,R_K, B)$ given a minimum energy-rate $b$ required at the EH are achievable at an $\eta$-NE. These $\eta$-NE sum-rate optimal tuples are Pareto optimal points of the information-energy capacity region $\set{E}(K,b)$.  
This suggests that, under the assumption that the players are able to properly choose the operating $\eta$-NE, for instance via learning algorithms, there is no loss of performance in the decentralized SIET case in comparison  to the fully centralized SIET case.

\subsection{Case of Two-Users: Example and Observations}
 
\begin{figure}[ht]
	\centering{
		\vspace*{-3mm}
		\includegraphics[scale=0.6]{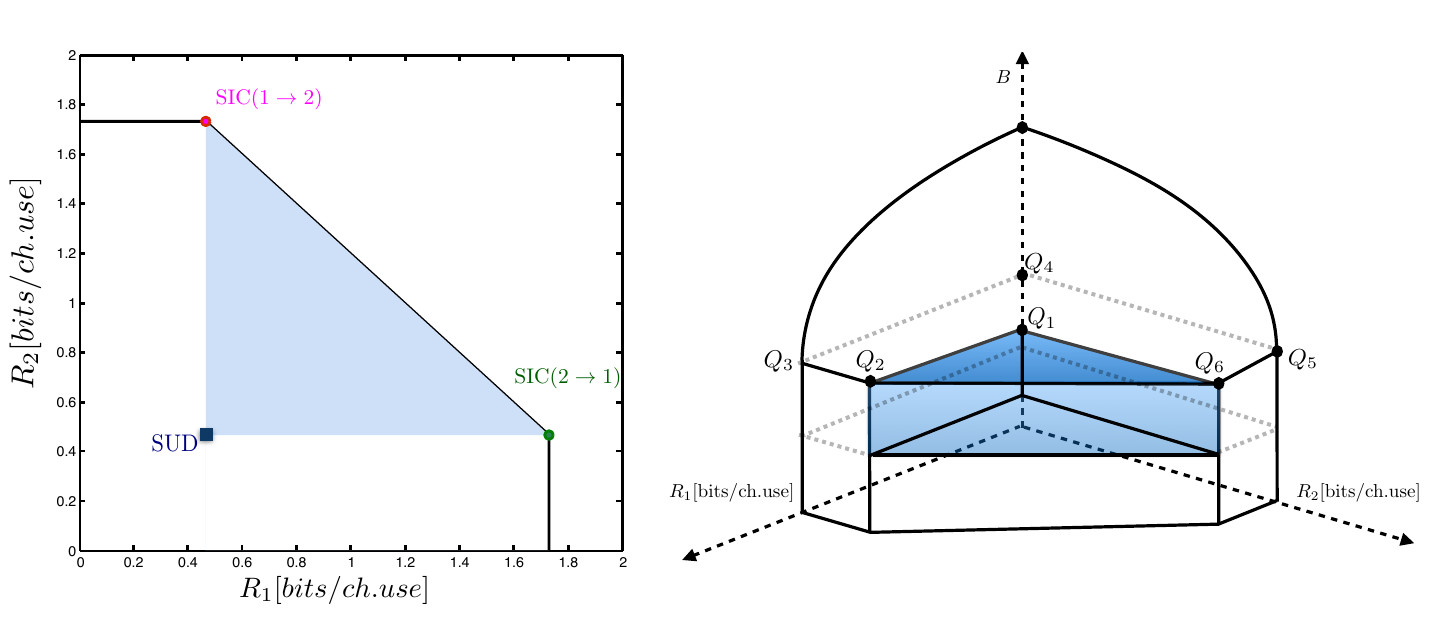}
		\vspace*{-5mm}
		\caption{Left figure depicts the the projection of the sets $\set{N}_{\mathrm{SUD}}(2,b)$ (square point),  $\set{N}_{\mathrm{SIC}(i\to j)}(2,b)$ (round points) (Decoding order: encoder $i$ before encoder $j$), and $\set{N}(b)$ (blue region) over the $R_1$-$R_2$ plane for $b\leqslant b_\ind(2)$. The information capacity region is also plotted as a reference (white region inside solid lines) for $\SNR_{11}=\SNR_{12}=\SNR_{21}=\SNR_{22}=10$. Note that the information capacity region with and without energy transmission rate constraint are identical in this case.
			Right figure is a 3-D representation of $\set{N}(b)$ (blue volume). The information-energy capacity region $\set{E}(2,0)$ is also plotted as a reference (white volume inside solid lines) for $\SNR_{11}=\SNR_{12}=\SNR_{21}=\SNR_{22}=10$.	
			\label{Fig2}}
	} 	
\end{figure}

\begin{figure}[ht]
	\centering{
		\includegraphics[scale=0.6]{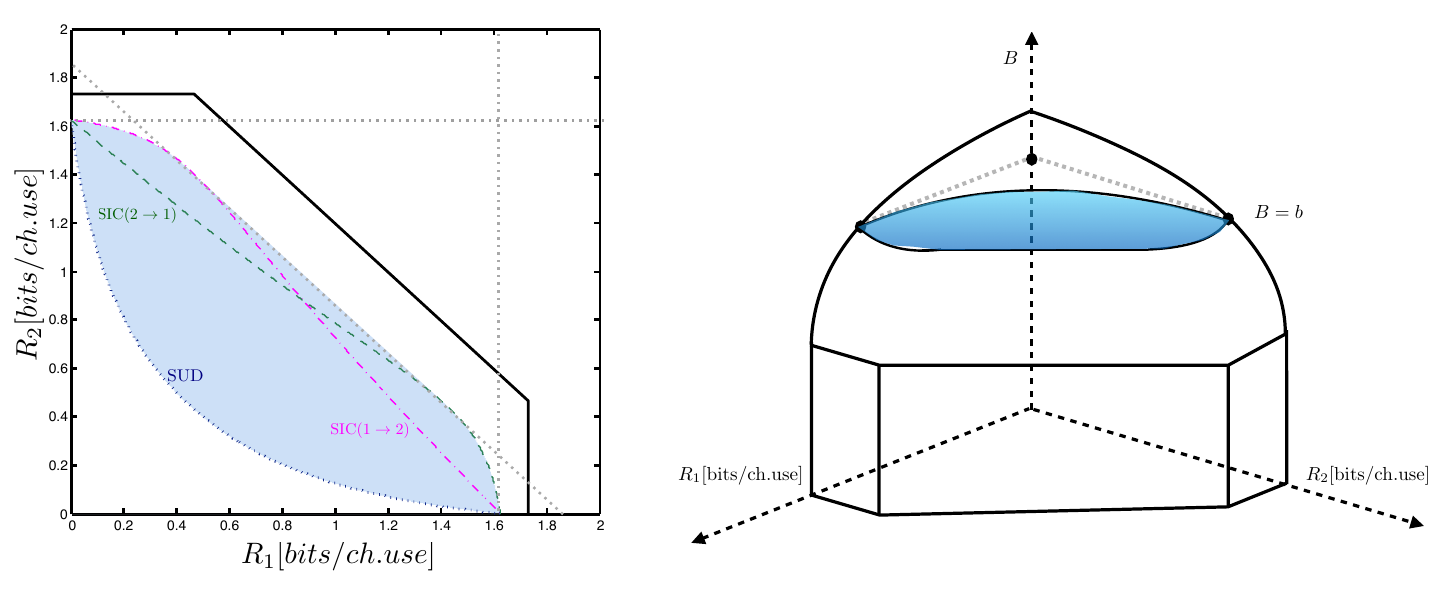}
		\vspace*{-5mm}
		\caption{Left figure depicts the projection of the sets $\set{N}_{\mathrm{SUD}}(b)$ (dotted line), $\set{N}_{\mathrm{SIC}(i \to j)}(b)$ (dashed lines) (Decoding order: encoder $i$ before encoder $j$), and $\set{N}(b)$ (blue region) over the $R_1$-$R_2$ plane for $b=0.7 B_{\max} > b_\ind(2)$. The information capacity region without energy transmission constraints (region inside solid lines) is plotted  for $\SNR_{11}=\SNR_{12}=\SNR_{21}=\SNR_{22}=10$ (Note that $B_{\max}\eqdef 1+\SNR_{21}+\SNR_{22}+2\sqrt{\SNR_{21}\SNR_{22}}$).
			Right figure is a 3-D representation of $\set{N}(b)$ (blue volume). The information-energy capacity region $\set{E}(2,0)$ is also plotted as a reference (white volume inside solid lines) for $\SNR_{11}=\SNR_{12}=\SNR_{21}=\SNR_{22}=10$.	 
			\label{Fig3}}
	}
\end{figure}

Consider the two-user symmetric G-MAC with $\SNR_{11} = \SNR_{12} = \SNR_{21} = \SNR_{22} = 10$ (EH and receiver are co-located). Note that for all $b \leqslant b_\ind(2)$,  the two transmitters use all their available average power for information transmission as shown in Fig.~\ref{Fig2}.
Alternatively, when $b > b_\ind(2)$, the two transmitters use the minimum energy needed to make the energy-outage probability arbitrarily close to zero and seek the largest possible information transmission rate (See Fig.~\ref{Fig3}). 

\section{Proof of Theorem~\ref{thm1}}
\label{sec:proof_thm1}
Consider the set of information-energy rate-tuples that can be achieved under the assumption that the receiver performs SUD to recover the messages $M_1,\dots, M_K$. This set is denoted by  $\set{C}_{\SUD}(K,b)$ and is defined by the following lemma. 
\begin{lemma}[Achievable Information-Energy Region With SUD]
	The set $\set{C}_{\SUD}(K,b)$ contains all non-negative information-energy rate tuples $(R_1,\dots,R_K,B)$ that satisfy
	\begin{subequations}
		\label{eq:ach_SUD}	
		\begin{IEEEeqnarray}{ccccl}
			0&\leqslant& R_i &\leqslant& \frac12 \log_2\left(1+\frac{\beta_i\SNR_{1i}}{1+\sum_{j=1;j\neq i}^K\beta_j\SNR_{1j}}\right),\quad i \in \{1,\dots,K\}, \label{eq:ach_SUDRi}\\
			b&\leqslant& B   & \leqslant & 1 + \sum_{j=1}^K \beta_j\SNR_{2j} + \left(  \sum_{j=1}^K   \sqrt{(1-\beta_j)\SNR_{2j}} \right)^2,
		\end{IEEEeqnarray}
	\end{subequations}		
	where $(\beta_1,\dots,\beta_K)\in [0,1]^K$.
\end{lemma}
\begin{IEEEproof}The proof of achievability follows similar arguments to those in  the proof of Theorem~\ref{TheoremCentralizedRegion} when the decoder is restricted to use SUD to recover the messages $M_1,\dots,M_K$. 
\end{IEEEproof} 	

Let the subset $\set{U}_{\SUD}(K,b)\subseteq \set{C}_{\SUD}(K,b)$ contain all information-energy rate tuples $(R_1,\dots,R_K,B)\in \set{C}_{\SUD}(K,b)$ satisfying 
\begin{subequations}	     
			\begin{IEEEeqnarray}{ccccl}
		0&\leqslant& R_i &\leqslant& \frac12 \log_2\left(1+\frac{\beta_i\SNR_{1i}}{1+\sum_{j=1;j\neq i}^K\beta_j\SNR_{1j}}\right),\quad i \in \{1,\dots,K\}, \\
		b&\leqslant& B   & \leqslant & 1 + \sum_{j=1}^K \beta_j\SNR_{2j} + \left(  \sum_{j=1}^K   \sqrt{(1-\beta_j)\SNR_{2j}} \right)^2,\label{eq:32c}
	\end{IEEEeqnarray}
\end{subequations}
where $\beta_1=\dots=\beta_K=1$ when $b\in [0,b_\ind(K)]$ and $(\beta_1,\dots,\beta_K)\in [0,1]^K$ is chosen to satisfy the following equality 
\begin{equation}
1 + \sum_{j=1}^K \beta_j\SNR_{2j} + \left(  \sum_{j=1}^K   \sqrt{(1-\beta_j)\SNR_{2j}} \right)^2=b,
\end{equation}
when $b\in (b_\ind(K),b_\coop(K) ]$.

Let also the subset $\set{V}_{\SUD}(K,b)\subseteq\set{C}_{\SUD}(K,b)$ be defined as $\set{V}_{\SUD}(K,b)\eqdef \set{C}_{\SUD}(K,b)\setminus \set{U}_{\SUD}(K,b)$. Note that for any $b\in [0,b_\coop(K)]$, the sets $\set{V}_{\SUD}(K,b)$ and $\set{U}_{\SUD}(K,b)$ form a partition of $\set{C}_{\SUD}(K,b)$.

To prove Theorem~\ref{thm1}, the first step is to show that
\begin{equation}
\set{N}_{\SUD}(K,b)\subseteq \set{U}_{\SUD}(K,b).
\end{equation}
That is, any achievable information-energy rate tuple $(R_1,\dots,R_K,B)\in \set{V}_{\SUD}(K,b)$ cannot be an $\eta$-NE with $\eta>0$, i.e., $\set{V}_\SUD(K,b)\cap \set{N}_\SUD(K,b)=\emptyset$. This is proved by Proposition~\ref{prop1SUD}.
\begin{proposition}
	\label{prop1SUD}
	Any information-energy rate tuple $(R_1,\dots,R_K,B)\in \set{V}_{\SUD}(K,b)$ 
	is not an $\eta$-NE, with $\eta>0$. That is, 
	\begin{equation}
	\set{N}_{\SUD}(K,b)\subseteq \set{U}_{\SUD}(K,b).\label{eq:34}
	\end{equation}
\end{proposition}

\begin{IEEEproof}The proof of Proposition~\ref{prop1SUD} is provided in Section~\ref{sec:proofprop1SUD}.
\end{IEEEproof}
The second step is to show that 
\begin{equation}
\set{U}_{\SUD}(K,b)\subseteq \set{N}_{\SUD}(K,b).
\end{equation}
That is, all information-energy rate tuples in $\set{U}_\SUD(K,b)$ are achievable for at least one $\eta$-NE, with $\eta>0$.
This is proved by Proposition~\ref{Prop2SUD}.
\begin{proposition}\label{Prop2SUD}
	Any information-energy rate tuple $(R_1,\dots,R_K,B)\in \set{U}_{\SUD}(K,b)$ is achievable at an $\eta$-NE, with an $\eta>0 $. That is,
	\begin{equation}
	\set{U}_{\SUD}(K,b)\subseteq \set{N}_{\SUD}(K,b).
	\end{equation}
\end{proposition}
\begin{IEEEproof}The proof of Proposition~\ref{Prop2SUD} is provided in Section~\ref{sec:proofprop2SUD}.
\end{IEEEproof}
This completes the proof of Theorem~\ref{thm1}.

\subsection{Proof of Proposition~\ref{prop1SUD}}
\label{sec:proofprop1SUD}

Any information-energy rate tuple $(R_1,\dots,R_K,B) \in  \set{V}_\SUD(K,b)$ satisfies at least one of the following conditions:
\begin{IEEEeqnarray}{rcl}
R_i &<&\frac12 \log_2\left(1+\frac{\beta_i\SNR_{1i}}{1+\sum_{j=1;j\neq i}^K\beta_j\SNR_{1j}}\right),\quad i \in \{1,\dots,K\},\label{eq:cond1}\\
B&<&b,\label{eq:cond4}\\
\label{eq:cond3}
B&>&1 + \sum_{j=1}^K \beta_j\SNR_{2j} + \left(  \sum_{j=1}^K   \sqrt{(1-\beta_j)\SNR_{2j}} \right)^2,
\end{IEEEeqnarray}
where $\beta_1=\dots=\beta_K=1$ when $b\in [0,b_\ind(K)]$ and $\beta_1,\dots,\beta_K$ are chosen to satisfy the following equality 
\begin{equation}
b=1 + \sum_{j=1}^K \beta_j\SNR_{2j} + \left(  \sum_{j=1}^K   \sqrt{(1-\beta_j)\SNR_{2j}} \right)^2,
\end{equation}
when $b\in (b_\ind(K), b_\coop(K) ]$.

Before introducing the proof of Proposition~\ref{prop1SUD},  some necessary conditions for $\eta$-NE action profiles are identified. These conditions are provided by Lemmas~\ref{lemCR} and~\ref{lemIIDG}. Under these necessary conditions, it is later shown that any rate tuple  $(R_1,\dots,R_K,B)$ that satisfies at least one of the conditions \eqref{eq:cond1}--\eqref{eq:cond3} is not an $\eta$-NE, with $\eta>0$. This establishes the proof of Proposition~\ref{prop1SUD}.

\subsubsection{Necessary Conditions on $\eta$-NE Actions}
Let $(R_1^*,\dots,R_K^*,B^*)$ be  an $\eta$-NE for any $\eta>0$, achievable by an action profile $(s_1^*,\dots,s_K^*)\in \set{A}_1\times\dots\times\set{A}_K$.

Denote by $X_{i,1}^*,\dots,X_{i,n}^*$ the channel inputs generated by transmitter~$i$, for all $i\in \{1,\dots,K\}$, with the equilibrium action $s_i^*$ and denote by $P_i^*$ their average power, that is
\begin{equation}
P_i^*\eqdef \frac1n \sum_{t=1}^n \E{(X_{i,t}^*)^2}.
\end{equation}

From the assumption that $(R_1^*,\dots,R_K^*,B^*)$ is achievable,  $P_{\error}^{(n)}(R_1^*,\dots,R_K^*)$ and $P_\outage^{(n)}(B^*)$ can be made arbitrarily small.
Thus, from \eqref{EqUtilityTX} it follows that
\begin{subequations}
	\begin{IEEEeqnarray}{lcl}
		u_i(s_1^*,\dots,s_K^*)&=&R_i^* , \quad\forall i \in \{1,\dots,K\}. 
	\end{IEEEeqnarray} 
\end{subequations}
Using this notation, the following lemma can be stated.
\begin{lemma}[Common Randomness]
	\label{lemCR}
	A necessary condition for the action profile $(s_1^*,\dots,s_K^*)$ to be an $\eta$-NE action is  that,  if the channel inputs $X_{i,t}^*$ are of the form $X_{i,t}^*=X_{i,1,t}^*+X_{i,2,t}^*$ where $X_{i,1,t}^*$ and $X_{i,2,t}^*$ are an information-carrying component and a non-information-carrying component, respectively,  then, $X_{i,2,t}^*$  must exclusively carry common randomness that is known to the receiver, for $i\in \{1,\dots,K\}$. 
\end{lemma}	
\begin{IEEEproof}
	Without loss of generality, consider transmitter $1$ whose utility is given by
	\begin{equation}
	u_1(s_1^*,s_2^*,\dots,s_K^*)=R_1^*.\label{eq:46}
	\end{equation}
	From the assumptions of the lemma, 
	component $X_{i,2,t}^*$ does not increase the information rate. Let $R_1$ denote the information rate that can be achieved by transmitter 1 if the interference caused by its component $X_{i,2,t}^*$ can be  completely canceled at the receiver before decoding the messages $M_1,\dots,M_K$.
	
	Assume that, in the action $s_1^*$, the component $X_{i,2,t}^*$ does not exclusively carry common randomness that is known to the receiver. 
	Hence, the receiver is not able to cancel the energy-carrying component before decoding it. This additional interference reduces the information rate of transmitter $1$. Let $\delta>0$ denote the penalty on the information rate of transmitter $1$ that is caused by this additional interference. That is,
	\begin{equation}
	R_1^*=R_1-\delta.\label{eq:47}
	\end{equation}
	
	Regardless of the actions $s_2^*,\dots,s_K^*$, transmitter~$1$ can use an alternative action $s_1$ in which the component $X_{i,2,t}^*$ carries only common randomness known to the receiver. Thus, its effect can be completely canceled and the information transmission can be performed at rate $R_1$.
	The corresponding utility is
	\begin{equation}
	u_1(s_1,s_2^*,\dots,s_K^*)=R_1.\label{eq:48}
	\end{equation} 
	From \eqref{eq:46}, \eqref{eq:47}, and \eqref{eq:48},  it holds that
	\begin{equation}
	u_1(s_1,s_2^*,\dots,s_K^*)-u_1(s_1^*,s_2^*,\dots,s_K^*)=R_1-R_1^*=\delta >0.
	\end{equation}
	The utility improvement is bounded away from zero, and thus the action profile $(s_1^*,\dots,s_K^*)$ cannot be an $\eta$-NE (Def.~\ref{DefEtaNE}), with an $\eta> 0$.
\end{IEEEproof}

\begin{remark}
	\label{rem1}
	Since the messages $M_1,\dots,M_K$ are independent, the only possible source of correlation between the\break time-$t$ channel inputs $X_{1,t},\dots,X_{K,t}$ is the common randomness that is known non-causally to all the transmitters and to the receiver. Furthermore, negatively correlating the inputs results in a loss of energy rate as well as information rates. Hence, a necessary condition for an $\eta$-NE is that the correlation must be non-negative.
\end{remark}

\begin{lemma}[IID Gaussian Inputs With Maximum Power]
	\label{lemIIDG}
	A necessary condition for the action profile $(s_1^*,\dots,s_K^*)$ to be an $\eta$-NE action is  that the input symbols $X_{i,t}^*$, with $i \in \{1,\dots,K\}$, are  generated i.i.d.~following a zero-mean Gaussian distribution with variance $P_i^*=P_{i,\max}$. 
\end{lemma}
\begin{IEEEproof} 
	Without loss of generality, consider transmitter $1$ and let $\tilde R_1$ denote the information rate that can be achieved by transmitter $1$ when the input symbols are generated i.i.d.~following a Gaussian distribution with maximum power $P_{1,\max}$ and where the information-carrying components of the transmitters are uncorrelated. 
	
	Assume that in the action $s_1^*$,  the input symbols are not generated i.i.d.\ following a Gaussian distribution with variance $P_1^*$.
	Since Gaussian distribution maximizes the entropy and since the information rates are increasing in the input power, using non-Gaussian inputs or using less power results in a loss in the achievable information rate. Thus, in the action $s_1^*$ the utility of transmitter 1 is 
	\begin{equation}\label{eq:35}
	u_1(s_1^*,s_2^*,\dots,s_K^*)=R_1^*=\tilde R_1-\zeta,
	\end{equation}
	where $\zeta>0$ quantifies the loss in information rate. 
	
	From the assumption that the receiver implements SUD,  independently of the actions $s_2^*,\dots,s_K^*$ of the other transmitters, there always exists an alternative action $s_1$ in which transmitter $1$ uses i.i.d.\ Gaussian codebooks with variance $P_1^*=P_{1,\max}$,  which achieves an information rate (and thus a utility) 
	\begin{equation}
	u_1(s_1,s_2^*,\dots,s_K^*)=\tilde R_1.\label{eq:36}
	\end{equation}
	
	From \eqref{eq:35} and \eqref{eq:36}, it follows that 
	\begin{IEEEeqnarray}{rCl}
		u_1(s_1,s_2^*,\dots,s_K^*)-u_1(s_1^*,s_2^*,\dots,s_K^*)=\zeta
		&>&0.
	\end{IEEEeqnarray}
	The utility improvement is bounded away from zero, and thus the action profile $(s_1^*,\dots,s_K^*)$ cannot be $\eta$-NE (Def.~\ref{DefEtaNE}), with an $\eta> 0$.
\end{IEEEproof}

\subsubsection{Proof of \eqref{eq:cond1}}
Without loss of generality, consider transmitter $1$ and consider the action profile $(s_1^*,\dots,s_K^*)$.

From Lemmas~\ref{lemCR} and \ref{lemIIDG}, a necessary condition for the action $s_1^*$ to be an $\eta$-NE action is to have i.i.d.~Gaussian channel inputs with maximum power $P_{1,\max}$ in which the energy-carrying component exclusively carries common randomness known to the receiver. This condition implies that any information rate $R_1$ satisfying 
\begin{equation}
0\leqslant R_1 \leqslant \frac12 \log_2\left(1+\frac{\beta_1 \SNR_{11}}{1+\sum_{j=2}^K\beta_j\SNR_{1j}}\right), 
\end{equation}
can be achieved with an arbitrarily small probability of error.

Assume that in the action profile $(s_1^*,\dots,s_K^*)$, the information rate $R_1^*$ satisfies
\begin{equation}
R_1^* <\frac12 \log_2\left(1+\frac{\beta_1\SNR_{11}}{1+\sum_{j=2}^K\beta_j\SNR_{1j}}\right), 
\end{equation}
and thus its utility satisfies
\begin{equation}
u_1(s_1^*,s_2^*,\dots,s_K^*)=R_1^*=\frac12 \log_2\left(1+\frac{\beta_1 \SNR_{11}}{1+\sum_{j=2}^K\beta_j\SNR_{1j}}\right)-\xi,\label{eq:56}
\end{equation} 
with $\xi>0$. 

Regardless of the action of the other transmitters, transmitter $1$ can always choose an alternative action $\tilde s_1$ in which it has a utility 
\begin{equation}
u_1(\tilde s_1,s_2^*,\dots,s_K^*)=\tilde R_1=\frac12 \log_2\left(1+\frac{\beta_1 \SNR_{11}}{1+\sum_{j=2}^K\beta_j\SNR_{1j}}\right).\label{eq:57}
\end{equation} 

From \eqref{eq:56} and \eqref{eq:57}, it holds that
\begin{equation}
u_1(\tilde s_1,s_2^*,\dots,s_K^*)-u_1(s_1^*,s_2^*,\dots,s_K^*)=\xi>0.
\end{equation} 
The utility improvement is bounded away from zero, and thus the action profile $(s_1^*,\dots,s_K^*)$ cannot be an $\eta$-NE (Def.~\ref{DefEtaNE}), with an $\eta> 0$.

The same reasoning holds for any transmitter $i$, with $i \in \{2,\dots,K\}$, and thus an action profile $(s_1^*,\dots,s_K^*)$ which induces an information-energy rate tuple $(R_1^*,\dots,R_K^*,B^*)$ for which $$R_i^* < \frac12 \log_2\left( 1+ \frac{\beta_i\SNR_{1i}}{1+\sum_{j=1;j\neq i}^K\beta_j\SNR_{1j}}\right)$$ for at least one $i\in \{1,\dots,K\}$,  cannot an $\eta$-NE, with $\eta > 0$.

\subsubsection{Proof of \eqref{eq:cond4}}
This is trivial since, $B^*< b$ implies that the tuple $(R_1^*,\dots,R_K^*,B^*)$ is not achievable (Def.~\ref{DefAchievableTriples}). 
\subsubsection{Proof of \eqref{eq:cond3}}

Assume that there exists an energy-information tuple $(R_1^*,\dots,R_K^*,B^*)$ that is achievable at an $\eta$-NE by the action profile $(s_1^*,\dots,s_K^*)\in \set{A}_1\times\dots\times\set{A}_K$ in which 
\begin{equation}
\label{eq:cond3B*}
B^*>1 + \sum_{j=1}^K \beta_j\SNR_{2j} + \left(  \sum_{j=1}^K   \sqrt{(1-\beta_j)\SNR_{2j}} \right)^2,
\end{equation}  
where $\beta_1=\dots=\beta_K=1$ when $b\in [0,b_\ind(K)]$ and $\beta_1,\dots,\beta_K$ are chosen to satisfy the following equality 
\begin{equation}
b=1 + \sum_{j=1}^K \beta_j\SNR_{2j} + \left(  \sum_{j=1}^K   \sqrt{(1-\beta_j)\SNR_{2j}} \right)^2,
\end{equation}
when $b\in (b_\ind(K),b_\coop(K) ]$. 
That is, \eqref{eq:cond3B*}
can equivalently be written as 
\begin{equation}
B^*>\max\{b,b_\ind(K)\}.\label{eq:61}
\end{equation}

From the previous parts of the proof (Lemma~\ref{lemCR} and Lemma~\ref{lemIIDG}), a necessary condition at any $\eta$-NE with $\eta> 0$ is to have the transmitters use Gaussian codebooks in which the channel inputs $\{(X_{1,t}^*,\dots,X_{K,t}^*)\}_{t=1}^n$ are generated i.i.d.~according to a Gaussian distribution with maximum powers $P_{1,\max},\dots,P_{K,\max}$, respectively. 

The fact that the energy rate cannot exceed the maximum feasible value given the constrained power budget at the transmitters, together with assumption~\eqref{eq:61} lead the following constraints on $B^*$: 
\begin{equation}
\max\{b,b_\ind(K)\} 
<B^*\leqslant b_\coop.\label{eq:60}
\end{equation}

Using continuity arguments,  the energy rate $B^*$ can be written as:
\begin{equation}
B^*=b_\ind(K)+2\sum_{1\leqslant i< j\leqslant K} \lambda_{ij} \sqrt{\SNR_{2i}\SNR_{2j}}
\end{equation}
where $0\leqslant \lambda_{ij} \leqslant 1$ denotes the Pearson correlation coefficient between $X_{i,t}$ and $X_{j,t}$ (See Remark~\ref{rem1}). 

Since the only source of correlation is common randomness whose effect is canceled before decoding the messages (See Remark~\ref{rem1}), for any values of the utililities $(R_1^*,\dots,R_K^*)$, one can always find $(\beta_1,\dots,\beta_K)$ satisfying 
\begin{equation}
\lambda_{ij}=\sqrt{(1-\beta_i)(1-\beta_j)}.
\end{equation} 
such that the utilities at the $\eta$-NE (Recall the necessary condition at any $\eta$-NE with $\eta>0$) can be written as 
\begin{IEEEeqnarray}{lcl}
	u_i(s_1^*,\dots,s_K^*)&=& R_i^*=\frac12 \log_2\left(1+\frac{\beta_i \SNR_{1i}}{1+\sum_{j=1;j\neq i}^K\beta_j \SNR_{1j}}\right)-\epsilon_i,\label{eq:util1}
\end{IEEEeqnarray}
with $\epsilon_i>0$ arbitrarily small.
One way to construct the channel inputs $X_{i,t}^*$ is 
\begin{equation}
X_{i,t}^*=\sqrt{(1-\beta_i) P_{i,\max}} U_t^*+ \sqrt{\beta_i P_{i,\max}} V_{i,t}^*,\quad 	i\in \{1,\dots,K\},
\end{equation}
where $U^*$,  $V_1^*,\dots,$ and $V_K^*$ are zero-mean unit-variance Gaussian RVs that are mutually independent. The variable $U^*$ depends on the common randomness $\Omega$ and the variable $V_i^*$ depends on the message $M_i$ for $i\in \{1,\dots,K\}$.

The strict inequality in~\eqref{eq:61} implies that for at least a pair of transmitters $k$ and $\ell$, the channel inputs $X_{k,t}^*$ and $X_{\ell,t}^*$ are positively correlated, i.e., $\lambda_{k\ell}>0$, which implies that $\beta_k<1$ and $\beta_\ell<1$.

For these two transmitters, the input correlation results in an information rate-loss and their utilities will be given by
\begin{IEEEeqnarray}{lcl}
	u_k(s_1^*,\dots,s_K^*)&=& R_k^*=\frac12 \log_2\left(1+\frac{\beta_k\SNR_{1k}}{1+\sum_{j=1;j\neq k}^K \beta_j\SNR_{1j}}\right)-\delta_k,\label{eq:utilk}\\
	u_\ell(s_1^*,\dots,s_K^*)&=& R_\ell^*=\frac12 \log_2\left(1+\frac{\beta_{\ell}\SNR_{1\ell}}{1+\sum_{j=1;j\neq \ell}^K \beta_j\SNR_{1j}}\right)-\delta_\ell,\label{eq:utilell}
\end{IEEEeqnarray} 
for some $\delta_k>0$ and $\delta_\ell>0$. 

Regardless of the actions of the other transmitters, transmitter $k$ can always use an alternative strategy $\tilde s_k$ in which it uses a power fraction $\beta_k'>\beta_k$. This reduces the correlation with the channel input of transmitter $\ell$ and increases the information rate of transmitter $k$ while keeping the energy rate above the threshold $b$. 

With the new strategy, transmitter $k$ achieves the information rate 
\begin{align}
\tilde R_k&=\frac12 \log_2\left(1+\frac{\beta_k'\SNR_{1k}}{1+\sum_{j=1;j\neq k}^K \beta_{j} \SNR_{1j}}\right),\\
\tilde B&\geqslant b.
\end{align}
Thus, the resulting utility of transmitter $k$ is 
\begin{IEEEeqnarray}{lcl}
	u_k(s_1,s_2^*,\dots,s_K^*)&=&\frac12 \log_2\left(1+\frac{\beta_k'\SNR_{1k}}{1+\sum_{j=1;j\neq k}^K \beta_{j} \SNR_{1j}}\right). \label{eq:71}
\end{IEEEeqnarray}

From \eqref{eq:util1} and \eqref{eq:71}, it follows that
\begin{IEEEeqnarray}{rCl}
	u_k(s_1,s_2^*,\dots,s_K^*)-u_k(s_1^*,s_2^*,\dots,s_K^*)&=&\frac12 \log_2\left(1+\frac{(\beta_k'-\beta_k)\SNR_{1k}}{1+\sum_{j=1;j\neq k}^K \beta_{j} \SNR_{1j}}\right)+\delta_k>0,
\end{IEEEeqnarray}
which contradicts the assumption that $(s_1^*,\dots,s_K^*)$ is an $\eta$-NE (Def.~\ref{DefEtaNE}), with an  $\eta> 0$.

This completes the proof of Proposition~\ref{prop1SUD}.
\subsection{Proof of Proposition~\ref{Prop2SUD}}	
\label{sec:proofprop2SUD}
Let $\eta > 0$ be fixed and assume that the decoder performs SUD. 

\textbf{Case 1:} $0\leqslant b\leqslant b_\ind(K)$:

Consider the rate tuple $(R_1^*,\dots,R_K^*,B^*)$ satisfying
\begin{subequations}
	\begin{IEEEeqnarray}{lcl}
		R_i^*&=&\frac12 \log_2\left(1+\frac{\SNR_{1i}}{1+\sum_{j=1;j\neq i}^K\SNR_{1j}}\right), \quad\forall i \in \{1,\dots,K\},\label{eq:68a}\\
		B^*&=& b_\ind(K).
	\end{IEEEeqnarray}
\end{subequations}
The targeted energy rate $b$ is less than what is strictly necessary to guarantee reliable communication at maximum information sum-rate. Thus,  the energy rate constraint is vacuous and the transmitters can exclusively use all their available power budget to send information, i.e., $\beta_1=\dots=\beta_K=1$. 

To achieve $(R_1^*,\dots,R_K^*,B^*)$,	transmitters $1,\dots,K$ can use the action profile $(s_1^*,\dots,s_K^*)$ described in the sequel.  The transmitters use independent Gaussian codebooks with powers $P_{1,\max},\dots,P_{K,\max}$,  as in~\cite{COVER75} or \cite{WYNER74}. The messages $M_1,\dots,M_K$ are encoded at the information rates $R_1^*,\dots,R_K^*$, respectively. The resulting average energy rate at the input of the EH is given by $B^{(n)}=b_\ind(K)$, which ensures that the energy outage probability $P_\outage^{(n)}(B^*)$ can be made arbitrarily small as the blocklength tends to infinity.		
From the assumption that the receiver performs SUD, the probability of error  $P_{\error}^{(n)}(R_1^*,\dots,R_K^*)$ can be made arbitrarily small as the blocklength tends to infinity.		
Hence, the resulting utilities are given by:
\begin{subequations}
	\begin{IEEEeqnarray}{lcl}
		u_i(s_1^*,s_2^*,\dots,s_K^*)&=&R_i^*,\quad \forall i \in \{1,\dots,K\}.\label{eq:69a} 
	\end{IEEEeqnarray} 
\end{subequations}

Assume that the action profile $(s_1^*,\dots,s_K^*)$ is not an $\eta$-NE. Then, from Def.~\ref{DefEtaNE}, there exist at least one player $i\in\{1,\dots,K\}$ and at least one strategy $\tilde s_i\in \set{A}_i$ such that the utility $u_i$ is improved by at least $\eta$ bits per channel use when player $i$ deviates from $s_i^*$ to $\tilde s_i$.  

Without loss of generality, let transmitter $1$ be the deviating player and denote by $\tilde R_1$ its new information rate. Hence,
\begin{equation}
\label{eq:70}
u_1(\tilde s_1,s_2^*,\dots,s_K^*)=\tilde R_1> u_1(s_1^*,s_2^*,\dots,s_K^*)+\eta. 
\end{equation}
From \eqref{eq:68a}, \eqref{eq:69a}, and \eqref{eq:70}, it holds that
\begin{equation}
\tilde R_1> \frac12 \log_2\left(1+\frac{\SNR_{11}}{1+\sum_{j=2}^K\SNR_{1j}}\right)+\eta.
\end{equation}
As the information rate tuple $(R_1^*,R_2^*,\dots,R_K^*)$ already saturates  the decoding capability of the receiver, the new information rate pair $(\tilde R_1,R_2^*,\dots, R_K^*)$ cannot be achieved and will result in a probability of error bounded away from zero. Consequently, the corresponding utility will be:
\begin{equation}
u_1(\tilde s_1,s_2^*,\dots, s_K^*)=-1,
\end{equation}
which contradicts the initial assumption \eqref{eq:70} and establishes that 
the action profile $(s_1^*,\dots,s_K^*)$ is an $\eta$-NE.
For the same information rates $(R_1^*,\dots,R_K^*)$, for any energy rate $B$, with $b\leqslant B \leqslant b_\ind(K)$, the information-energy rate tuple $(R_1^*,\dots,R_K^*,B)$ is also achievable by the same action profile $(s_1^*,\dots,s_K^*)$. Note that $(R_1^*,\dots,R_K^*,B)$ is also achievable at an $\eta$-NE.

\textbf{Case 2:} $b_\ind(K)< b\leqslant b_\coop(K)$:

Consider the information-energy rate tuple $(R_1^*,\dots,R_K^*,B^*)$ such that:
\begin{subequations}	
	\begin{IEEEeqnarray}{lcl}
		R_i^* &=& \frac12 \log_2\left(1+\frac{\beta_i^*\SNR_{1i}}{1+\sum_{j=1;j\neq i}\beta_j^*\SNR_{1j}}\right),\; \forall i \in \{1,\dots,K\},\\
		B^* &= & 1 +\sum_{j=1}^K \beta_j^*\SNR_{2j} + \left(  \sum_{j=1}^K   \sqrt{(1-\beta_j^*)\SNR_{2j}} \right)^2,
	\end{IEEEeqnarray}
\end{subequations}
where  $(\beta_1^*,\dots,\beta_K^*)$ are chosen to satisfy 
\begin{equation}
b= 1 +\sum_{j=1}^K \beta_j^*\SNR_{2j} + \left(  \sum_{j=1}^K   \sqrt{(1-\beta_j^*)\SNR_{2j}} \right)^2.
\end{equation}

To achieve $(R_1^*,\dots,R_K^*,B^*)$,	transmitters $1$ to $K$ can use the action profile $(s_1^*,\dots,s_K^*)$ described in the sequel.  Transmitters $1$ to $K$ use power fractions $\beta_1^*,\dots,\beta_K^*$ of their power budgets $P_{1,\max},\dots,P_{K,\max}$ to send information using independent Gaussian codebooks as in~\cite{COVER75,WYNER74}, 
and use the remaining power ($(1-\beta_1^*) P_{1,\max},\dots,(1-\beta_K^*)P_{K,\max}$) to send common randomness  that is known to all the transmitters and the receiver. This common randomness does not carry any information and does not produce any interference to the information-carrying signals. The messages $M_1,\dots,M_K$ are encoded at the information rates $R_1^*,\dots,R_K^*$ chosen by transmitters $1$ to $K$, respectively. The receiver first subtracts the common randomness and then performs SUD to recover the messages $M_1,\dots,M_K$.

The resulting average energy rate at the input of the EH is given by $B^{(n)}= 1+\sum_{j=1}^K \beta_j^*\SNR_{2j} + \left(  \sum_{j=1}^K   \sqrt{(1-\beta_j^*)\SNR_{2j}} \right)^2$. This ensures that $B^*\geqslant b$ and that the energy outage probability $P_\outage^{(n)}(B^*)$ can be made arbitrarily small as the blocklength tends to infinity.		

Assume that the action profile $(s_1^*,\dots,s_K^*)$ is not an $\eta$-NE. Then, from Def.~\ref{DefEtaNE}, there exist at least one player $i\in\{1,\dots,K\}$ and at least one strategy $\tilde s_i\in \set{A}_i$ such that the utility $u_i$ is improved by at least $\eta$ bits per channel use when player $i$ deviates from $s_i^*$ to $\tilde s_i$. 

Without loss of generality, let transmitter $1$ be the deviating player and denote by $\tilde R_1$ its new information rate. Hence,
\begin{equation}
\label{eq:75}
u_1(\tilde s_1,s_2^*,\dots,s_K^*)=\tilde R_1> u_1(s_1^*,s_2^*,\dots,s_K^*)+\eta. 
\end{equation}

The new information-energy rate tuple $(\tilde R_1,R_2^*,\dots, R_K^*,B^*)$ is outside the information-energy capacity region and will result in a utility 
\begin{equation}
u_1(\tilde s_1,s_2^*,\dots,s_K^*)=-1 
\end{equation}
which contradicts the assumption \eqref{eq:75} and establishes that $(s_1^*,\dots,s_K^*)$ is an $\eta$-NE.

\section{Proof of Theorem~\ref{thm2}}
\label{sec:proof_thm2}
The proof of Theorem~\ref{thm2} follows along the same lines as the proof of Theorem~\ref{thm1} when considering the set of information-energy rate tuples which can be achieved if the receiver performs $\SIC(\pi)$, for a fixed decoding order $\pi \in \set{P}_K$ to recover the messages $M_1,\dots,M_K$. This set is denoted by  $\set{C}_{\SIC(\pi)}(b)$ and is defined as the set of $(R_1,\dots,R_K,B)$ that satisfy
\begin{subequations}
	\label{eq:ach_SICij}	
	\begin{align}
		R_{\pi(i)} &= \frac12 \log_2\left(1+\frac{\beta_{\pi(i)}\SNR_{1\pi(i)}}{1+\sum_{j=i+1}^K \beta_{\pi(j)}\SNR_{1\pi(j)}}\right), \quad \forall i \in \{1,\dots,K\},\\
		b &\leqslant  B   \leqslant   1 + \sum_{j=1}^K \beta_j\SNR_{2j} + \left(  \sum_{j=1}^K   \sqrt{(1-\beta_j)\SNR_{2j}} \right)^2,
	\end{align}
\end{subequations}		
with $(\beta_1,\dots,\beta_K)\in [0,1]^K$ are feasible power-splits, i.e., 
\begin{equation}
b\leqslant 1 + \sum_{j=1}^K \beta_j\SNR_{2j} + \left(  \sum_{j=1}^K   \sqrt{(1-\beta_j)\SNR_{2j}} \right)^2.
\end{equation}
\section{Conclusion}
\label{sec:conc}
In this paper, the fundamental limits of decentralized SIET in the $K\geqslant 2$-user G-MAC with minimum received energy rate constraint have been derived in terms of $\eta$-NE regions, with $\eta> 0$. A key observation in this work is the fact that the decentralization induces no loss of performance for SIET as long as the players are able to properly choose the operating $\eta$-NE for instance via learning algorithms. Recently, Belhadj Amor \textit{et al.} have shown that channel output feedback enhances SIET as it provides additional cooperation among the users. An interesting open question is whether feedback may help in the decentralized case.
 Also, the knowledge given to each player and the order in which actions can be taken substantially change the nature of the game  and the corresponding stable region. Furthermore, such a region varies if a different notion of equilibrium is considered, e.g., Stackelberg equilibrium~\cite{Stackelberg}, correlated equilibrium~\cite{Aumann}, satisfaction equilibrium~\cite{Perlaza-JSSP2012}, etc.

\balance
\bibliographystyle{IEEEtran}

\begin{thebibliography}{10}
	\providecommand{\url}[1]{#1}
	\csname url@samestyle\endcsname
	\providecommand{\newblock}{\relax}
	\providecommand{\bibinfo}[2]{#2}
	\providecommand{\BIBentrySTDinterwordspacing}{\spaceskip=0pt\relax}
	\providecommand{\BIBentryALTinterwordstretchfactor}{4}
	\providecommand{\BIBentryALTinterwordspacing}{\spaceskip=\fontdimen2\font plus
		\BIBentryALTinterwordstretchfactor\fontdimen3\font minus
		\fontdimen4\font\relax}
	\providecommand{\BIBforeignlanguage}[2]{{%
			\expandafter\ifx\csname l@#1\endcsname\relax
			\typeout{** WARNING: IEEEtran.bst: No hyphenation pattern has been}%
			\typeout{** loaded for the language `#1'. Using the pattern for}%
			\typeout{** the default language instead.}%
			\else
			\language=\csname l@#1\endcsname
			\fi
			#2}}
	\providecommand{\BIBdecl}{\relax}
	\BIBdecl
	
	\bibitem{Tesla1914-Patent}
	\BIBentryALTinterwordspacing
	N.~Tesla, ``Apparatus for transmitting electrical energy,'' Dec. 1914, {US}
	Patent 1119732. [Online]. Available:
	\url{https://www.google.com/patents/US1119732}
	\BIBentrySTDinterwordspacing
	
	\bibitem{Wheeler-1943}
	L.~P. Wheeler, ``{T}esla's contribution to high frequency,'' \emph{Electrical
		Engineering}, vol.~62, no.~8, pp. 355--357, Aug. 1943.
	
	\bibitem{BelhadjAmor-ICT-2016}
	S.~{Belhadj Amor} and S.~M. Perlaza, ``Fundamental limits of simultaneous
	energy and information transmission,'' in \emph{Proc.~23rd International
		Conference on Telecommunications}, Thessaloniki, Greece, May 2016.
	
	\bibitem{VAR12}
	L.~R. Varshney, ``On energy/information cross-layer architectures,'' in
	\emph{Proc.~IEEE International Symposium on Information Theory}, Jul. 2012,
	pp. 1356--1360.
	
	\bibitem{Fouladgar-CL-2012}
	A.~M. Fouladgar; and O.~Simeone, ``On the transfer of information and energy in
	multi-user systems,'' \emph{IEEE Communications Letters}, vol.~16, no.~11,
	pp. 1733--1736, Nov. 2012.
	
	\bibitem{BelhadjAmor-TIT-2017}
	S.~{Belhadj Amor}, S.~M. Perlaza, I.~Krikidis, and H.~V. Poor, ``Feedback
	enhances simultaneous information and energy transmission in multiple access
	channels,'' \emph{IEEE Transactions on Information Theory}, vol.~63, no.~8,
	pp. 5244--5265, Aug. 2017.
	
	\bibitem{BelhadjAmor-COMNET-2015}
	S.~{Belhadj Amor}, S.~M. Perlaza, and I.~Krikidis, ``Simultaneous energy and
	information transmission in {Gaussian} multiple access channels,'' in
	\emph{Proc.~5th International Conference on Communications and Networking
		(ComNet)}, Hammamet, Tunisia, Nov. 2015.
	
	\bibitem{Khalfet-Globecom}
	N.~Khalfet and S.~M. Perlaza, ``Simultaneous information and energy
	transmission in {G}aussian interference channels,'' submitted to the 2018
	International Zurich Seminar on Information and Communication, Zurich, Feb.
	2018.
	
	\bibitem{Khalfet-Allerton}
	------, ``Simultaneous information and energy transmission in {G}aussian
	interference channels with feedback,'' in \emph{Proc.~55th Annual Allerton
		Conference on Communication, Control, and Computing}, Monticello, IL, USA,
	Oct. 2017.
	
	\bibitem{Nash-PNAS-1950}
	J.~F. Nash, ``Equilibrium points in $n$-person games,'' \emph{Proc. of the
		National Academy of Sciences}, vol.~36, pp. 48--49, Jan. 1950.
	
	\bibitem{Stackelberg}
	V.~H. Stackelberg, \emph{Marketform und Gleichgewicht}.\hskip 1em plus 0.5em
	minus 0.4em\relax Oxford University Press, 1934.
	
	\bibitem{Aumann}
	R.~J. Aumann, ``Subjectivity and correlation in randomized strategies,''
	\emph{Journal of Mathematical Economics}, vol.~1, no.~1, pp. 67--96, Mar.
	1974.
	
	\bibitem{Perlaza-JSSP2012}
	S.~M. Perlaza, H.~Tembine, S.~Lasaulce, and M.~Debbah, ``Quality-of-service
	provisioning in decentralized networks: A satisfaction equilibrium
	approach,'' \emph{IEEE Journal of Selected Topics in Signal Processing},
	vol.~6, no.~2, pp. 104--116, Apr. 2012.
	
	\bibitem{Lai}
	L.~Lai and H.~{El Gamal}, ``The water-filling game in fading multiple-access
	channels,'' \emph{IEEE Transactions on Information Theory}, vol.~54, no.~5,
	pp. 2110--2122, May 2008.
	
	\bibitem{GajicRimoldi}
	V.~Gajic and B.~Rimoldi, ``Game theoretic considerations for the {Gaussian}
	multiple access channel,'' in \emph{Proc.~IEEE International Symposium on
		Information Theory}, Toronto, ON, Canada, Jul. 2008, pp. 2523--2527.
	
	\bibitem{Netgcoop-2016}
	S.~{Belhadj Amor} and S.~M. Perlaza, ``Decentralized $k$-user {G}aussian
	multiple access channels,'' in \emph{Proc.~International conference on
		NETwork Games, Control and OPtimization (NETGCOOP 2016)}, Avignon, France,
	Nov. 2016, pp. 45--55.
	
	\bibitem{Breton-JOA-1988}
	M.~Breton, A.~Alj, and A.~Haurie, ``Sequential {S}tackelberg equilibria in
	two-person games,'' \emph{Journal of Optimization Theory and Applications},
	vol.~59, no.~1, pp. 37--43, Oct. 1988.
	
	\bibitem{Varan}
	B.~Varan and A.~Yener, ``Incentivizing signal and energy cooperation in
	wireless networks,'' \emph{IEEE Journal on Selected Areas in Communications},
	vol.~33, no.~12, pp. 2554--2566, Dec. 2015.
	
	\bibitem{COVER75}
	T.~M. Cover, \emph{Some Advances in Broadcast Channels}.\hskip 1em plus 0.5em
	minus 0.4em\relax Academic Press, 1975, vol.~4, ch.~4.
	
	\bibitem{WYNER74}
	A.~D. Wyner, ``Recent results in the {S}hannon theory,'' \emph{IEEE
		Transactions on Information Theory}, vol.~20, no.~1, pp. 2--10, Jan. 1974.
	
	\bibitem{Nisan-Book-2007}
	N.~Nisan, T.~Roughgarden, E.~Tardos, and V.~V. Vazirani, \emph{Algorithmic Game
		Theory}.\hskip 1em plus 0.5em minus 0.4em\relax New York, USA: Cambridge
	University Press, 2007.
	
\end{thebibliography}

\end{document}